\RequirePackage{cmap}

\RequirePackage[x11names,dvipsnames,svgnames]{xcolor}

\documentclass[]{aa}
\makeatletter
\renewcommand*\aa@pageof{, page \thepage{} of \pageref*{LastPage}}
\makeatother

\usepackage{silence}
\usepackage[T1]{fontenc}
\usepackage{textcomp}

\usepackage{amsmath}
\usepackage{amssymb}
\usepackage{bm}
\usepackage{bbm}

\usepackage{fontawesome5}

\usepackage[varg]{txfonts}
\usepackage[scaled=0.76]{beramono}

\usepackage[inline]{enumitem}

\usepackage{nicefrac}

\usepackage{csquotes}

\usepackage{microtype}

\usepackage{siunitx}
\DeclareSIUnit\bar{bar}
\DeclareSIUnit\parsec{pc}
\DeclareSIUnit\year{a}

\usepackage{flushend}

\usepackage{chemformula}

\usepackage{dcolumn} %

\usepackage[
    pdftex,
    unicode=true,
    colorlinks,
    urlcolor=Blue,
    linkcolor=gray,
    citecolor=gray,
    filecolor=gray,
    pdfauthor={},
    plainpages=false,
    pdfpagelabels,
]{hyperref}

\usepackage[noabbrev, capitalise]{cleveref}

\usepackage{url}

\usepackage[all]{hypcap}
\usepackage{natbib}
\bibpunct{(}{)}{;}{a}{}{,}

\let\oldbibliography\thebibliography
\renewcommand{\thebibliography}[1]{%
  \oldbibliography{#1}%
  \setlength{\itemsep}{2.5pt}%
  \setlength{\baselineskip}{7.5pt}
  \setlength{\lineskiplimit}{-\maxdimen}
}

\usepackage{standalone}

\usepackage{graphicx}

\usepackage{subcaption}

\usepackage{rotating}
\usepackage{pdflscape}

\usepackage{placeonpage}

\usepackage{booktabs}
\usepackage{threeparttable} %

\definecolor{C0}{HTML}{1f77b4}
\definecolor{C1}{HTML}{ff7f0e}
\definecolor{C2}{HTML}{2ca02c}

\definecolor{CBF0}{HTML}{5790FC}
\definecolor{CBF1}{HTML}{F89C20}
\definecolor{CBF2}{HTML}{E42536}
\definecolor{CBF3}{HTML}{964A8B}
\definecolor{CBF4}{HTML}{9C9CA1}
\definecolor{CBF5}{HTML}{7A21DD}

\usepackage{tikz}
\usetikzlibrary{
    decorations.pathreplacing,
    matrix,
    patterns,
    positioning,
    shapes.geometric,    
}

\usepackage{xspace}

\newcommand{\given}{\ensuremath{\,|\,}}

\DeclareMathOperator*{\argmin}{arg\,min}

\newcommand{\R}{\ensuremath{\mathbb{R}}}
\newcommand{\z}{\ensuremath{\bm{z}}}

\newcommand{\pyatmos}{\textsc{PyATMOS}\xspace}
\newcommand{\goyal}{\textsc{Goyal-2020}\xspace}

\newcommand{\codelink}[1]{\href{#1}{\scalebox{0.75}{\faIcon[regular]{file-code}}}}

\usepackage[precision=2]{lengthconvert}

\newcommand{\labelbox}[4]{%
    \begin{tikzpicture}
        \draw [draw=none, fill=none] (0,0) rectangle (#1 - \pgflinewidth, #2 - \pgflinewidth) node[pos=0.5, align=center, font=\tiny\bfseries\sffamily, rotate=#3] {#4};
    \end{tikzpicture}%
}

\newenvironment{notes}[1]
{%
  \par\addvspace{6pt}
  \begin{spacing}{0.81}
  \tiny
  \noindent \textit{#1.} \hskip.5em
}
{%
  \end{spacing}
} 

\begin{document}

\title{
    Parameterizing pressure-temperature profiles\\ of exoplanet atmospheres with neural networks
}
\titlerunning{
    Parameterizing pressure-temperature profiles of exoplanet atmospheres with neural networks
}

\author{
    Timothy D. Gebhard\inst{1,2,3}\fnmsep%
    \thanks{Correspondence: \href{mailto:tgebhard@tue.mpg.de}{tgebhard@tue.mpg.de}.} \and %
    Daniel Angerhausen\inst{3} \and %
    Björn S. Konrad\inst{3} \and \\%
    Eleonora Alei\inst{3} \and %
    Sascha P. Quanz\inst{3} \and %
    Bernhard Schölkopf\inst{1,4} %
}
\authorrunning{Gebhard et al. (2023)}

\institute{
    Max Planck Institute for Intelligent Systems, Max-Planck-Ring 4, 72076 Tübingen, Germany \and %
    Max Planck ETH Center for Learning Systems, Max-Planck-Ring 4, 72076 Tübingen, Germany \and %
    ETH Zurich, Institute for Particle Physics \& Astrophysics, Wolfgang-Pauli-Str. 27, 8092 Zurich, Switzerland \and %
    Department of Computer Science, ETH Zurich, 8092 Zurich, Switzerland
}

\date{Version: \today}

\abstract{
    Atmospheric retrievals (AR) of exoplanets typically rely on a combination of a Bayesian inference technique and a forward simulator to estimate atmospheric properties from an observed spectrum. 
    A key component in simulating spectra is the pressure-temperature (PT) profile, which describes the thermal structure of the atmosphere.
    Current AR pipelines commonly use ad hoc fitting functions here that limit the retrieved PT profiles to simple approximations, but still use a relatively large number of parameters.
}{
    In this work, we introduce a conceptually new, data-driven parameterization scheme for physically consistent PT profiles that does not require explicit assumptions about the functional form of the PT profiles and uses fewer parameters than existing methods.
}{
    Our approach consists of a latent variable model (based on a neural network) that learns a distribution over functions (PT profiles).
    Each profile is represented by a low-dimensional vector that can be used to condition a decoder network that maps $P$ to $T$.
}{
    When training and evaluating our method on two publicly available datasets of self-consistent PT profiles, we find that our method achieves, on average, better fit quality than existing baseline methods, despite using fewer parameters.
    In an AR based on existing literature, our model (using two parameters) produces a tighter, more accurate posterior for the PT profile than the five-parameter polynomial baseline, while also speeding up the retrieval by more than a factor of three.
}{
    By providing parametric access to physically consistent PT profiles, and by reducing the number of parameters required to describe a PT profile (thereby reducing computational cost or freeing resources for additional parameters of interest), our method can help improve AR and thus our understanding of exoplanet atmospheres and their habitability.
}

\keywords{
    methods: data analysis /
    methods: statistical /
    planets and satellites: atmospheres
}
     \maketitle

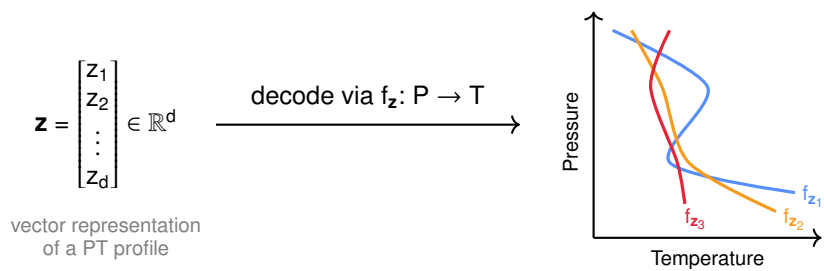
\begin{figure*}[t]
    \placeonpage{2}
    \centering
    \begin{tikzpicture}[font=\sffamily]
    
    \tikzset{every node/.append style={node font=\sffamily}}
    \tikzstyle{arrow}=[->];

    \node [inner sep=0] (z) {\textbf{z} = $\begin{bmatrix}
           \text{z\textsubscript{1}} \\
           \text{z\textsubscript{2}} \\
           \vdots \\
           \text{z\textsubscript{d}} \\
         \end{bmatrix}$ $\in \mathbb{R}^\text{d}$ };
    \node [below=2mm of z, font=\tiny, align=center, gray] {vector representation\\ of a PT profile};

    \node [right=5cm of z] (cs) {};

    \draw [arrow, thick, shorten >= 5mm, shorten <= 5mm] (z.east) -- node [pos=0.5, above=1mm, font=\normalsize] {decode via f\textsubscript{\textbf{z}}: P $\to$ T} (cs);

    \begin{scope}[shift=(cs), xshift=0.35cm, yshift=-1.5cm]
    
    \draw [arrow, thick] (0, 0) -- node [rotate=90, anchor=center, pos=0.65, left=3mm, font=\tiny, inner sep=0] {Pressure} (0, 3);
    \draw [arrow, thick] (0, 0) -- node [pos=0.5, inner sep=0, below=1.5mm, font=\tiny] {Temperature} (3, 0);

    \draw [draw=CBF0, very thick] plot [smooth, tension=0.5] coordinates { (0.25, 2.75) (1.50, 2.00) (1.00, 1.00) (2.65, 0.60)};
    \node [font=\tiny, CBF0, inner sep=0] at (2.75+0.15, 0.55) {f\textsubscript{\textbf{z}\textsubscript{1}}};
    
    \draw [draw=CBF1, very thick] plot [smooth, tension=0.5] coordinates { (0.5, 2.75) (0.90, 2.00) (1.25, 1.00) (2.4, 0.35)};
    \node [font=\tiny, CBF1, inner sep=0] at (2.5+0.15, 0.35-0.1) {f\textsubscript{\textbf{z}\textsubscript{2}}};
    
    \draw [draw=CBF2, very thick] plot [smooth, tension=0.5] coordinates { (1.0, 2.75) (0.75, 2.00) (1.1, 1.00) (1.2, 0.45)};
    \node [font=\tiny, CBF2, inner sep=0] at (1.2+0.15, 0.35-0.1) {f\textsubscript{\textbf{z}\textsubscript{3}}};
        
    \end{scope}
    
\end{tikzpicture}
    \caption{
        Schematic illustration of the problem setting considered in this paper:
        \enquote{Parameterizing PT profiles} means finding a way to represent PT profiles as vectors $\z \in \R^d$, together with a decoding mechanism $f_{\z}: P \to T$ that converts these $\z$-vectors back into a function that maps pressure values onto temperature values.
        Our goal in this work is to learn both the representations $\z$ and the mechanism $f_{\z}$ from data obtained with simulations using full climate models.
    }
    \label{fig:problem}
\end{figure*}

\begin{figure*}[t]
    \placeonpage{3}
    \centering
    \begin{tikzpicture}[font=\sffamily]

    % Define colorblind-friendly colors
    \definecolor{C0}{HTML}{5790fc}  % blue
    \definecolor{C1}{HTML}{f89c20}  % orange
    \definecolor{C2}{HTML}{e42536}  % red
    \definecolor{C3}{HTML}{964a8b}  % purple
    \definecolor{C4}{HTML}{9c9ca1}  % gray
    \definecolor{C5}{HTML}{7a21dd}  % purple
    
    % Define \rvdots command
    \makeatletter
    \DeclareRobustCommand{\rvdots}{%
      \vbox{
        \baselineskip4\p@\lineskiplimit\z@
        \kern-\p@
        \hbox{.}\hbox{.}\hbox{.}
      }}
    \makeatother

    \tikzset{every node/.append style={node font=\sffamily}}

    \tikzstyle{arrow}=[thick, ->];
    \tikzstyle{square}=[regular polygon, regular polygon sides=4];
    \tikzstyle{box}=[square, minimum width=9mm, draw, thick, inner sep=0.5mm, rounded corners, align=center, font=\tiny];
    \tikzstyle{C0}=[draw=C0, text=C0];
    \tikzstyle{C1}=[draw=C1, text=C1];
    \tikzstyle{C2}=[draw=C2, text=C2];

    % Dots
    \node [box, draw=none] (p_dots_E) {\rvdots};
    \node [left=1mm of p_dots_E, box, draw=none] (t_dots_E) {\rvdots};

    % Encoder inputs
    \node [above=1mm of p_dots_E, box, C0] (p_1) {p\textsubscript{1}};
    \node [left=1mm of p_1, box, C1] (t_1) {t\textsubscript{1}};
    \draw [thick] (p_1) -- (t_1);
    \draw [arrow] (p_1) -- ++(0.825, 0);
    \node [below=1mm of p_dots_E, box, C0] (p_N) {p\textsubscript{N}};
    \node [left=1mm of p_N, box, C1] (t_N) {t\textsubscript{N}};
    \draw [thick] (p_N) -- (t_N);
    \draw [arrow] (p_N) -- ++(0.825, 0);
    \coordinate (x) at (p_N.west)+(-0.05, 0);
    \node [below=4.5mm of x, font=\tiny, align=center, gray] {Input PT profile};

    % Encoder
    \node [right=5mm of p_dots_E, trapezium, minimum height=8mm, trapezium angle=130, align=center, shape border rotate=45, draw, thick] (E) {E};
    
    % Latent code
    \node [right=25mm of E, circle, draw, thick, minimum height=6.5mm, inner sep=0, C2] (z) {\textbf{z}};
    \node [below=0.5mm of z, font=\tiny, align=center, gray] {(Latent) representation\\ of the input PT profile.\\[1mm] Should follow a given\\ prior distribution.};

    \draw [draw=none, fill=none]  (z) rectangle ++(9.2, 2);
    \draw [draw=none, fill=none]  (z) rectangle ++(9.2, -2);
    \draw [draw=none, fill=none] (z) rectangle ++(-9.2, 2);
    \draw [draw=none, fill=none] (z) rectangle ++(-9.2, -2);

    % Decoders
    \node [right=15mm of z, trapezium, minimum height=6.5mm, trapezium angle=75, align=center, shape border rotate=0, draw, thick] (D_1) {D};
    \node [right=17.5mm of D_1, trapezium, minimum height=6.5mm, trapezium angle=75, align=center, shape border rotate=0, draw, thick] (D_d) {D};

    % Decoder inputs
    \node [above=6.5mm of D_1, box, C0] (p_1_in) {p\textsubscript{1}};
    \node [above=6.5mm of D_d, box, C0] (p_d_in) {p\textsubscript{N}};

    % Decoder outputs
    \node [below=6.5mm of D_1, box, C1] (t_1_out) {$\hat{\text{t}}$\textsubscript{1}};
    \node [below=6.5mm of D_d, box, C1] (t_d_out) {$\hat{\text{t}}$\textsubscript{N}};

    % Dots
    \node [right=2.5mm of p_1_in, trapezium, minimum height=6.5mm, trapezium angle=75, align=center, shape border rotate=0] (p_in_dots) {$\cdots$};
    \node [below=6.5mm of p_in_dots, trapezium, minimum height=6.5mm, trapezium angle=75, align=center, shape border rotate=0] (D_dots) {$\cdots$};
    \node [right=2.5mm of t_1_out, trapezium, minimum height=6.5mm, trapezium angle=75, align=center, shape border rotate=0] (t_out_dots) {$\cdots$};
   
    \draw [] (z.east) edge[arrow, bend left=30, C2] (D_1.west);
    \draw [] (z.east) edge[arrow, bend left=30, C2] node [pos=0.12, above=1.5mm, inner sep=0, rotate=22.5, align=center, font=\tiny, C2, draw=none] {condition\\ D on z}  (D_d.west);

    \draw [] (p_1_in) edge[arrow] (D_1.north);
    \draw [] (p_d_in) edge[arrow] (D_d.north);

    \draw [] (D_1.south) edge[arrow] (t_1_out.north);
    \draw [] (D_d.south) edge[arrow] (t_d_out.north);

    % Box around decoder
    \node [dotted, semithick, rounded corners=2.5mm, minimum width=72mm, minimum height=42mm, draw, anchor=west, xshift=-11.5mm] (decoder_box) at (z.west) {};
    \node [font=\tiny, fill=white] at (decoder_box.north) {Used during training and during a retrieval};
    
    % Box around encoder
    \node [dotted, semithick, rounded corners=2.5mm, minimum width=41.5mm, minimum height=42mm, draw, anchor=west, xshift=-2.5mm] (encoder_box) at (t_dots_E.west) {};
    \node [font=\tiny, fill=white] at (encoder_box.north) {Only used during training};

    % Arrow from encoder to z
    \draw [draw=white, line width=2.7mm] (E.east) -- (z.west);
    \draw [] (E.east) edge[arrow] node [pos=0.25, above=1.5mm, inner sep=0, align=center, font=\tiny] {encode} (z.west);

    % Add legend
    \node [right=16mm of D_dots, align=left, font=\tiny] {
        E:\\[1.5mm]
        D:\\[1.5mm]
        \textbf{z}:\\[1.5mm]
        p\textsubscript{i}:\\[1.5mm]
        t\textsubscript{i}:\\\\[1.5mm]
        $\hat{\text{t}}$\textsubscript{i}:\\
    };
        \node [right=20mm of D_dots, align=left, font=\tiny] {
        encoder network\\[1.5mm]
        decoder network\\[1.5mm]
        latent representation\\[1.5mm]
        pressure in \emph{i}-th layer\\[1.5mm]
        true temperature\\ in \emph{i}-th layer\\[1.5mm]
        predicted temperature\\ in \emph{i}-th layer
    };

\end{tikzpicture}
    \caption{
        Schematic illustration of our model:
        During training, the encoder network $E$ maps a PT profile (consisting of a vector $\bm{p} = (p_1, ..., p_N)$ of pressure values and a vector $\bm{t} = (t_1, ..., t_N)$ of corresponding temperature values) onto a latent representation $\z \in \R^d$.
        This latent code is then used to condition a decoder network $D$, and $D(\cdot \given \z)$ is evaluated on each $p_i$ to get the corresponding predicted temperatures $\hat{t}_i$.
        During an atmospheric retrieval, $\z$ is proposed by the Bayesian inference method (e.g., nested sampling).
    }
    \label{fig:our-method}
\end{figure*}
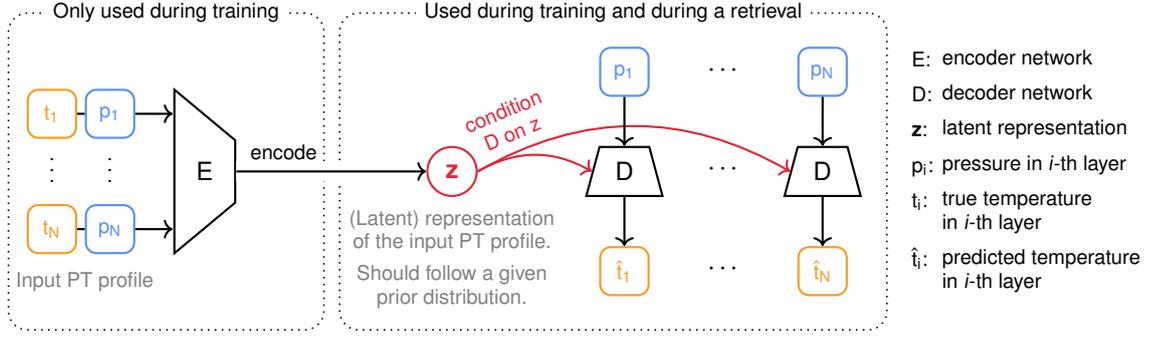

\begin{figure*}[t]
    \placeonpage{4}
    \centering
    \begin{subfigure}[t]{88mm}
        \centering
        \includegraphics[scale=1]{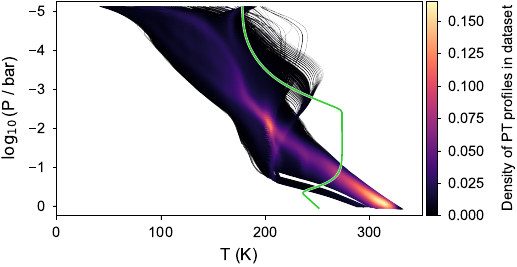}%
        \subcaption{\pyatmos}
        \label{fig:pyatmos-heatmap}
    \end{subfigure}%
    \hfill%
    \begin{subfigure}[t]{88mm}
        \centering
        \includegraphics[scale=1]{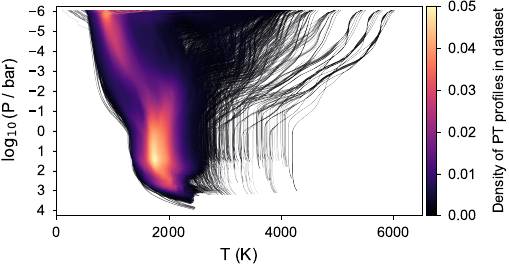}%
        \subcaption{\goyal}
        \label{fig:goyal-heatmap}
    \end{subfigure}%
    \caption{
        These plots show all the PT profiles in our two datasets.
        Each segment of a line---that is, the connection between the points $(p_i, t_i)$ and $(p_{i+1}, t_{i+1})$---is color-coded by the density of the profiles at its respective $(p, t)$ coordinate, which was obtained through a 2D KDE.
        The green line in the \pyatmos plot shows the one PT profile that we manually removed for being out-of-distribution.~%
        \codelink{https://github.com/timothygebhard/ml4ptp/blob/main/scripts/plotting/fig-3-datasets/plot-dataset.py}
    }
    \label{fig:heatmaps}
\end{figure*}

\begin{figure*}[t]
    \placeonpage{5}
    \centering
    \labelbox{2.5mm}{2.5mm}{0}{}\hfill%
    \labelbox{40mm}{2.5mm}{0}{\hspace{8mm}dim(z) = 1}\hfill%
    \labelbox{40mm}{2.5mm}{0}{\hspace{8mm}dim(z) = 2}\hfill%
    \labelbox{40mm}{2.5mm}{0}{\hspace{8mm}dim(z) = 3}\hfill%
    \labelbox{40mm}{2.5mm}{0}{\hspace{8mm}dim(z) = 4}\\
    \labelbox{2.5mm}{36mm}{90}{\hspace{6mm}Best RMSE}\hfill%
    \includegraphics{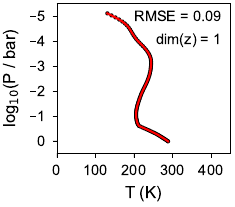}\hfill%
    \includegraphics{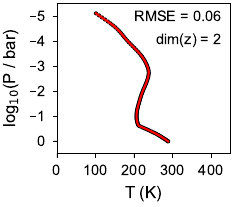}\hfill%
    \includegraphics{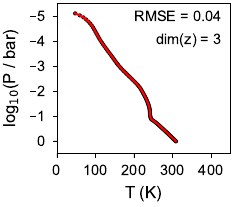}\hfill%
    \includegraphics{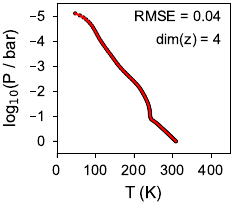}\\[2.5mm]
    \labelbox{2.5mm}{36mm}{90}{\hspace{6mm}Median RMSE}\hfill%
    \includegraphics{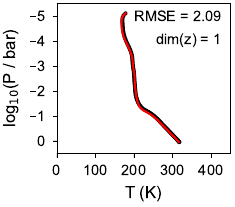}\hfill%
    \includegraphics{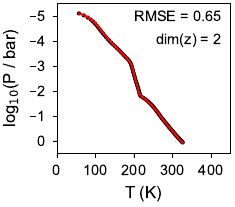}\hfill%
    \includegraphics{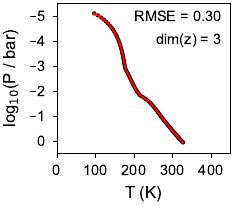}\hfill%
    \includegraphics{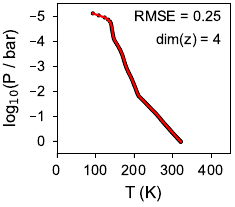}\\[2.5mm]
    \labelbox{2.5mm}{36mm}{90}{\hspace{6mm}Worst RMSE}\hfill%
    \includegraphics{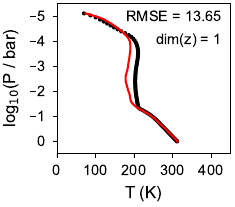}\hfill%
    \includegraphics{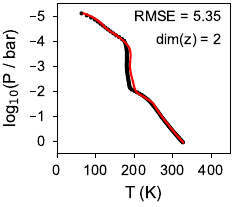}\hfill%
    \includegraphics{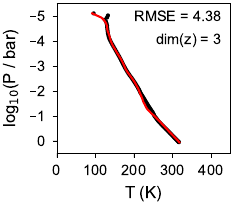}\hfill%
    \includegraphics{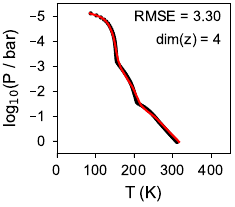}
    \caption{
        Examples of pressure-temperature profiles from our test set (black) together with the best approximation using our trained model (red).
        Columns show different values of $\dim(\z) \in \lbrace 1, 2, 3, 4 \rbrace$; rows show the best, median, and worst example (in terms of the RMSE).
        Results are shown for the \pyatmos dataset; additional plots for the \goyal are found in \cref{fig:example-profiles-goyal-2020} in the appendix.~%
        \codelink{https://github.com/timothygebhard/ml4ptp/blob/main/scripts/plotting/fig-4-example-pt-profiles/plot-pt-profile.py}
    }
    \label{fig:example-profiles-pyatmos}
\end{figure*}

\begin{figure*}[t]
    \placeonpage{6}
    \centering
    \includegraphics[]{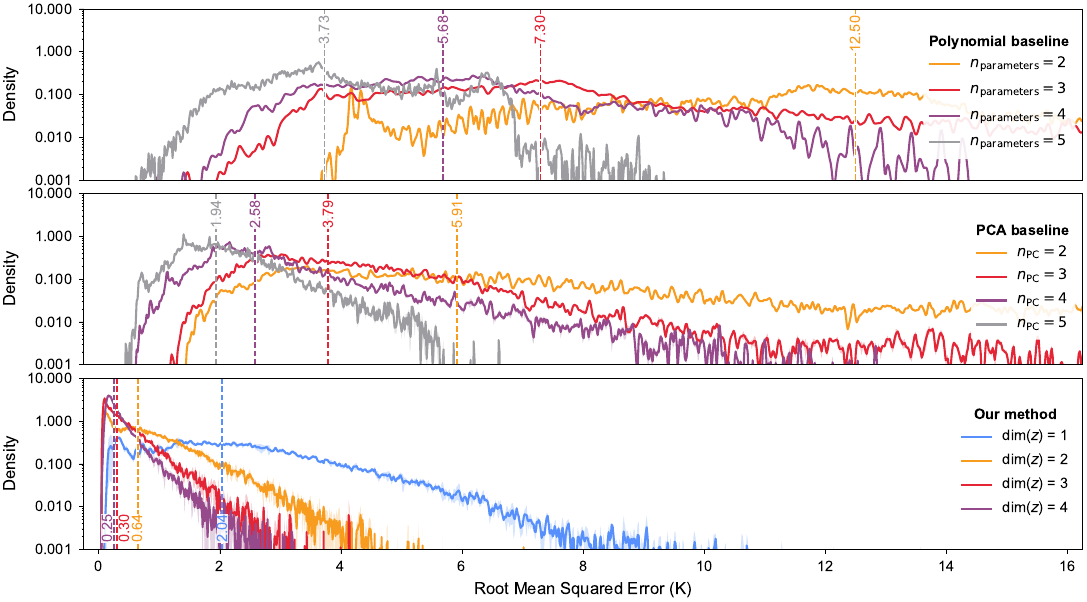}    
    \caption{
        Distributions of the reconstruction error (RMSE) for a polynomial baseline, a PCA-based baseline, and our method, for different number of fitting parameters.
        For our model, as well as the PCA baseline, each distribution is the average of three runs using different random seeds controlling the initialization and the train/validation split. 
        The dashed lines indicate the respective medians. 
        Results are shown for the \pyatmos dataset; additional plots for the \goyal are found in \cref{fig:error-distribution-goyal-2020} in the appendix.~%
        \codelink{https://github.com/timothygebhard/ml4ptp/blob/main/scripts/plotting/fig-5-error-distributions/plot-error-distributions.py}
    }
    \label{fig:error-distribution-pyatmos}
\end{figure*}

\begin{figure*}[t]
    \placeonpage{7}
    \centering
        \begin{subfigure}[t]{90mm}
        \centering
        \includegraphics{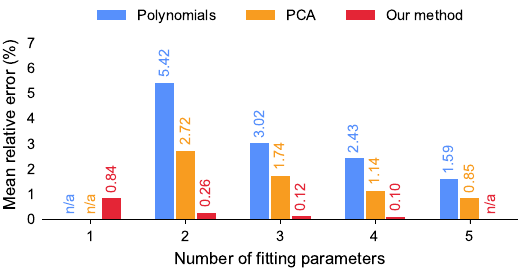}%
        \subcaption{PyATMOS}
    \end{subfigure}%
    \hfill%
    \begin{subfigure}[t]{90mm}
        \centering
        \includegraphics{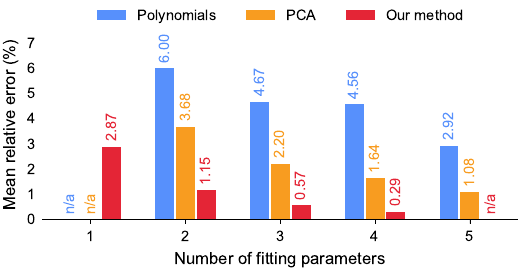}%
        \subcaption{Goyal-2020}
    \end{subfigure}%
    \caption{
        Comparison of the median test set MRE for the different methods and datasets (lower is better).
        Reminder: For each profile, we compute the \emph{mean} relative error (over all atmospheric layers), and then aggregate by computing the \emph{median} over all profiles.
        Finally, for the PCA baseline as well as our method, the results are also \emph{mean}-averaged over different random seeds.~%
        \codelink{https://github.com/timothygebhard/ml4ptp/blob/main/scripts/plotting/fig-6-mre-comparison/plot-mre-comparison.py}
    }
    \label{fig:mre-comparison}
\end{figure*}

\begin{figure*}
    \centering
    \placeonpage{8}
    \includestandalone{figures/fig-7-overview/fig-7-overview}
    \caption{
        A visual illustration of the encoder $E$, the decoder $D$, and their connection via the latent space. 
        See \cref{subsec:interpreting-latent-representations} in the main text for a detailed explanation.
        This figure is best viewed in color.~%
        \codelink{https://github.com/timothygebhard/ml4ptp/blob/main/scripts/plotting/fig-7-overview/create-plots.py}
    }
    \label{fig:latent-space}
\end{figure*}

\section{Introduction}

With now over \num{5000} confirmed planet detections outside the solar system~\citep[see, e.g.,][]{Christiansen_2022}, the exoplanet science community is increasingly expanding from detecting new planets to also characterizing them. 
One important tool for this are atmospheric retrievals~(AR), that is, \enquote{the inference of atmospheric properties of an exoplanet given an observed spectrum}~\citep{Madhusudhan_2018}.
The properties of interest here include, for example, the chemical composition of the atmosphere (i.e., abundances of different chemical species) or the presence of clouds or haze.
Deriving these properties from a spectrum is a classic example of an inverse problem. 
As with many such problems, the standard approach is to combine Bayesian inference methods, such as nested sampling \citep{Skilling_2006}, with a simulator for the \enquote{forward} direction---in our case, turning a set of atmospheric parameters $\bm{\theta}$ into the corresponding spectrum.
Given the spectrum $y$ of a planet of interest, one can then estimate a posterior distribution $p(\bm{\theta} \given y)$ by iteratively drawing a sample of parameter values from a prior $\pi(\bm{\theta})$, passing them to a simulator, and comparing the simulated spectrum $\hat{y}$ to the observed one through a likelihood function $L$ to guide the next parameter sample that is drawn.
In practice, the simulator is often deterministic, in which case $L$ is essentially equivalent to the noise model: 
For example, optimizing $\bm{\theta}$ to minimize $\lVert y - \hat{y}(\bm{\theta}) \rVert^2$ implies the assumption that the noise in $y$ comes from an uncorrelated Gaussian.

Traditionally, due to computational limitations, much of the existing work on atmospheric retrievals has made the simplifying assumption of treating atmospheres as one-dimensional. 
More recent studies, however, have also begun to explore 2D and 3D approaches \citep[see, e.g.,][]{Chubb_2022,Nixon_2022,Zingales_2022}.

\paragraph{Pressure-temperature profiles}
One of the key factors that determines the observable spectrum of a planet's atmosphere is its thermal structure, that is, the temperature of each atmospheric layer.
The thermal structure is described by the pressure-temperature profile, or PT profile, which is a function $f: \R^+ \to \R^+$ that maps an atmospheric pressure to its corresponding temperature.
In reality, of course, this function also depends on the location (e.g., in the case of the Earth, the air over the equator is warmer than over the poles), and such variations can be accounted for when using general circulation models (GCMs).
However, as mentioned above, it is common to limit oneself to one-dimensional approximations.
The implications of such an approximation are discussed, for example, in \citet{Blecic_2017}.

There are, in principle, two different ways to determine the PT profile during an AR:
One may either solve a set of (simplified) radiative transfer equations, or approximate the PT profile through an ad hoc fitting function~\citep[see, e.g.,][]{Seager_2010, Line_2013, Heng_2017}.
In the latter case, which is the one that we consider in this work, one has to choose a parameterization for the PT profile, that is, one needs to assume a parametric family of functions $f_{\z}: P \to T$ which map pressure values onto corresponding temperature values and have tunable parameters $\z \in \R^d$ that control how exactly this mapping looks like (see \cref{fig:problem} for an illustration).
These parameters $\z$ are a subset of the full set of parameters $\bm{\theta}$ for which the retrieval procedure estimates a posterior distribution (see above).
Throughout this work, we commonly refer to $\z$ as a (latent) representation of a PT profile.
The choice of $f_{\z}$ will, in general, be determined by a trade-off between different (and possibly conflicting) requirements and desiderata, including the following:

\begin{itemize}[leftmargin=*]
    \setlength\itemsep{0em}
    \item For every profile that we can expect to find during the retrieval, there should exist a $\z$ such that $f_{\z}$ is a good approximation of this profile (surjectivity).
    \item For every value of~$\z$ from some given domain (e.g., a subset of $\R^d$), $f_{\z}$ is a valid PT profile.
        Together with surjectivity, this implies that the image of $\z \mapsto f_{\z}$ is equal to the set of physically sensible (and relevant) PT profiles.
    \item The mapping $\z \mapsto f_{\z}$ should be smooth in the sense that small changes in $\z$ should only result in small changes in $f_{\z}$ (i.e., continuity); otherwise, the retrieval loop may not converge.
    \item The dimensionality $d = \dim(\z)$ should be as small as possible, since methods such as nested sampling often do not scale favorably in the number of retrieval parameters.
    \item We should be able to define a simple prior $p(\z)$ for $\z$, and for $\z \sim p(\z)$, the induced distribution over functions $f_{\z, \z \sim p(\z)}$ should constitute an appropriate prior distribution over the PT profiles of the population of planets that we are considering.
    \item Ideally, the dimensions of $\z$ are interpretable as physical quantities, or correspond to physical processes (e.g., \enquote{$z_1$ is the mean temperature,} or \enquote{$z_2$ controls the presence of an inversion}).
\end{itemize}

Of course, as suggested above, no choice of parameterization will perfectly satisfy all requirements simultaneously, and some sort of compromise is required.

\paragraph{Related work}
\citet{Barstow_2020}, who have identified the parameterization of PT profiles as one of the \enquote{outstanding challenges of exoplanet atmospheric retrievals,} find the approaches of \citet{Madhusudhan_2009} and \citet{Guillot_2010} to be the most popular among the atmospheric retrieval community.
These parameterizations, as well as the ones proposed by \citet{Hubeny_2003}, \citet{Hansen_2008}, \citet{Heng_2011}, and \citet{Robinson_2012}, all consist of (semi)-analytic expressions that seek to describe the relationship between pressure and temperature mainly in terms of physically interpretable quantities, such as the opacity or optical depth.
Simpler (yet less physically motivated) approaches include the use of splines~\citep[e.g.,][]{Zhang_2021} as well as low-order polynomials~\citep[e.g.,][]{Konrad_2022}. 
Finally, a data-driven approach was presented by \citet{Schreier_2020}, who compute a singular value decomposition of a representative set of PT profiles, and parameterize new profiles as a combination of the first $k$ singular vectors.
All these approaches typically use four to six parameters to describe a PT profile, which is often a substantial fraction of the total number of retrieved parameters.
To the best of our knowledge, no studies to date have used machine learning (especially neural networks) to parameterize pressure-temperature profiles of exoplanet atmospheres.

\paragraph{Contributions}
In this work, we explore a conceptually new approach to parameterizing PT profiles that is based on the idea of using neural networks to learn efficient (i.e., low-dimensional) representations of PT profiles, as well as the corresponding decoding mechanisms, from simulated data.%
\footnote{We presented early versions of this work at \mbox{AbSciCon~2022} and the \enquote{ML and the Physical Sciences} workshop at NeurIPS~2022.}
Using two different datasets of PT profiles obtained with full climate models, we show that our approach can achieve better fit quality with fewer parameters than our two baseline methods, while still easily integrating into an existing atmospheric retrieval framework.

\section{Method}
\label{sec:method}

Our goal is the following:
We want to learn a model (in our case: a neural network) that takes two inputs: (1) a pressure value $p$, and (2) a vector $\z \in \R^d$, also called \enquote{representation}, which contains a highly compressed description of a PT profile.
The output of the model is the temperature $t$ at pressure $p$, for the profile described by $\z$.
During an atmospheric retrieval, the Bayesian inference routine can then propose values of $\z$, which can be converted to a PT profile by conditioning the network on the proposed $\z$ (that is, fixing the input $\z$) and evaluating it at different pressures $p_i$ to get the corresponding temperatures $t_i$.
Thus, we will infer a posterior distribution over $\z$, which corresponds directly to a posterior over PT profiles.

To learn such a network, we assume that we have a dataset $\mathcal{D}$ of PT profiles which were generated by a full climate model or radiative transfer code to ensure they are physically consistent.
These models typically simulate an atmosphere in a layer-by-layer fashion, meaning that each PT profile is given in the form of a set support points, $\lbrace (p_i, t_i) \rbrace_{i=1,\ldots,N}$, where $N$ is the number of layers, and $p_i$ and $t_i$ are the pressure and temperature in the $i$-th layer.
Importantly, our method does not require that $p_i$ be the same for all profiles in $\mathcal{D}$: Different profiles can use different pressure grids, and this is indeed the case for the datasets we work with.
Moreover, in principle, even the number of layers $N$ can vary between profiles.
(We do not explicitly consider this case here, but it implies only minor technical complications at training time.)
This flexibility also makes it easy, for example, to combine PT profiles from different models or pre-existing atmosphere grids into a single training dataset.

The approach that we propose in the following is inspired by the idea of a (conditional) neural process~(NP; \citealt{Garnelo_2018a,Garnelo_2018b}).
A neural process is a type of machine learning model that learns a distribution over functions and combines the advantages of Gaussian processes with the flexibility and scalability of neural networks.
In astrophysics, the usage of (conditional) neural processes is still relatively rare; examples include \citet{Park_2021}, \citet{CvorovicHajdinjak_2021}, and \citet{Jankov_2022}, all of which use conditional NPs for interpolation.

Our method works as follows (see also \cref{fig:our-method}).
We employ two neural networks: an encoder $E$ and a decoder $D$.
The encoder is only required during training and will not be used during an AR.
At training time, we take a PT profile---consisting of a vector $\bm{p} = (p_1, ..., p_N)$ of pressure values and a vector $\bm{t} = (t_1, ..., t_N)$ of corresponding temperatures---from our training data set and pass it to $E$, which outputs a representation $\z \in \R^d$ of the profile:
\begin{equation}
    \z = E(\bm{p}, \bm{t}) \,.
\end{equation}
The dimensionality $d$ of $\z$ can be chosen freely (but changing $d$ will require re-training $E$ and $D$).
Generally, we want to keep $d$ as low as possible (e.g., $d = 2$), but of course, higher values also imply more expressive or informative representations.

In the next step, the representation $\z$ is then used to condition the decoder network $D$.
Once conditioned on $\z$, the decoder network \emph{is} a PT profile, that is, a function that takes in a single value (a pressure) and outputs a single value (a temperature).
We then pass all pressure values $p_i$ of our input profile through $D$ to get the respective predicted temperature values $\hat{t}_i$:
\begin{equation}
    \hat{t}_i = D(p_i \given \z) \,,
\end{equation}
where we use $D(\cdot \given \z): \R^+ \to \R^+$ to denote the function that is given by conditioning $D$ on $\z$  (i.e., fixing $\z$ as an input).
We train the encoder and decoder jointly (i.e., we update the weights of the neural networks) to minimize the difference between the true and predicted temperatures.
More specifically, we minimize the following mean squared error (MSE) reconstruction loss, which is a standard choice in machine learning for regression problems:
\begin{equation}
    \mathcal{L}_\text{rec}(\bm{t}, \bm{\hat{t}}) 
    = \frac{1}{N} \sum_{i=1}^{N} (t_i - \hat{t}_i)^2 \,.
    \label{eq:mse}
\end{equation}
In practice, we do this not only for a single PT profile, but minimize the average of $\mathcal{L}_\text{rec}$ over our entire training dataset in a batched fashion.
We also note that this choice of reconstruction loss assigns the same weight to all $N$ atmospheric layers, and refer to \cref{subsec:future-directions} for a brief discussion of possible alternatives.

To ensure that we can define a simple prior for $\z$ during a retrieval, we furthermore want to constrain the distribution of the $\z$'s produced by the encoder.
There are various potential approaches for this (see \cref{app:method}). 
For this work, we adopt an idea from \citet{Zhao_2017} and add a second term to the loss function that encourages $\z$ to follow a standard Gaussian:
\begin{equation}
    \mathcal{L}_\text{MMD} 
    = \text{MMD}^2\left( \lbrace \z_i \rbrace_{i=1,\ldots,b}, \lbrace \bm{s}_i \rbrace_{i=1,\ldots,b} \right) \,,
\end{equation}
where:
\begin{equation*}
    \bm{s}_i \sim \mathcal{N}_d(0, 1)\,, \enspace i=1, \ldots, b \,.
\end{equation*}
Here, $\mathcal{N}_d(0, 1)$ denotes a $d$-dimensional standard Gaussian distribution (i.e., mean zero and identity matrix as covariance matrix).
Furthermore, $b$ is the batch size of our training, and MMD stands for Maximum Mean Discrepancy \citep{Borgwardt_2006,Gretton_2012}, which is a kernel-based metric that measures whether two samples come from the same distribution (see \cref{subsec:mmd} for more details).
By minimizing this MMD term, we encourage the model to produce values of $\z$ that follow the same distribution as the $\bm{s}_i$, that is, a $d$-dimensional standard Gaussian.

Experimentally, we found that despite this MMD term, there are sometimes a few profiles that are mapped to a $\z$ very far from the origin, which is undesirable if we want to define a simple prior for $\z$ during a retrieval.
To prevent this problem, we introduce a third loss term with a softplus barrier function on the norm of $\z$:
\begin{align}
    \mathcal{L}_\text{norm}(\z) 
    &= \nicefrac{1}{k} \cdot \log \left( 1 + \exp \lbrace k \cdot (\lVert \z \rVert_2 - \tau) \rbrace \right) \,.
\end{align}
This is a smooth version of the $\mathrm{ReLU}(x) := \max(0, x)$ function, where $k$ controls the amount of smoothing:
Larger values of $k$ increase the similarity to ReLU.
For this work, we have fixed $k$ at $100$ (ad hoc choice), after observing that using a standard ReLU function destabilized the training for us.
The parameter $\tau$ defines the threshold on $\lVert \z \rVert_2$ and was set to $\tau = 3.5$ (ad hoc choice).

The total loss that we minimize is then just a weighted sum (with $\beta, \gamma \in \R^+$) of our three loss terms:
\begin{equation}
    \mathcal{L} = \mathcal{L}_\text{rec} + \beta \cdot \mathcal{L}_\text{MMD} + \gamma \cdot \mathcal{L}_\text{norm} \,.
\end{equation}

The hyperparameters $\beta$ and $\gamma$ control the tradeoff between the different loss terms:
(1)~Increasing $\beta$ corresponds to placing a stronger prior on the distribution of $\z$ in the latent space, usually at the cost of a higher reconstruction error.
We have tried different values of $\beta$ and found that the performance is relatively similar over a range of about 0.1 to 1.
Too high values of $\beta$ can lead to model collapse, where the prior on $\z$ becomes too strong and prevents the encoder from learning meaningful representations.
For this work, we set $\beta = 0.1$.
This was an ad hoc choice and not the result of systematic optimization, which would likely yield a different value for each dataset and value of $\dim(\z)$.
(2)~The value of $\gamma$ should be chosen such that $\gamma \cdot \mathcal{L}_\text{norm}$ dominates the other loss terms when $\lVert \z \rVert_2 > t$.
We set $\gamma = 10$ for this work (again without optimization).

To use our trained models during atmospheric retrieval, we provide a convenient Python wrapper that allows the model to be loaded as a normal function, taking as input a value for $\z$ and a \texttt{numpy} array of pressure values, and returning an array of corresponding temperatures.
We note again that neither the lengths of the pressure grids nor the exact values need to match the pressure grid(s) used during training.
(However, the range of pressure values should be approximately the same; otherwise, evaluating $D(\cdot \given \z)$ at pressure values it has never encountered during training may result in nonphysical outputs).

\section{Datasets}
\label{sec:datasets}

We validate our proposed approach using two different, publicly available datasets of PT profiles: one for terrestrial planets, and one for hot Jupiters.
A visual comparison of the two datasets and the respective $P$-$T$ space that they cover is shown in \cref{fig:heatmaps}.

\subsection{\pyatmos}
The \pyatmos dataset was generated by \citet{Chopra_2023} and consists of \num{124314} stable atmospheres of Earth-like planets around solar-type stars generated with \texttt{Atmos}, a 1D coupled photochemistry-climate model \citep{Arney_2016, Meadows_2018}.
The atmospheres in this dataset were simulated using $N = 101$ layers covering an altitude range of \SIrange{0}{80}{\kilo\meter}.
Consequently, each PT profile is given by two 101-dimensional vectors for the pressures and temperatures.
In addition, about 80 other atmospheric quantities are available, such as fluxes and abundance profiles of chemical species.
We manually removed one PT profile from the dataset because it was completely out-of-distribution and caused stability problems during training.

Furthermore, it has been brought to our attention that the simulations of the \pyatmos dataset suffered from a couple of issues that affect the physical correctness of the simulated PT profiles.
In particular, the \href{https://github.com/PyAtmos/atmos_fdl/blob/b2294838aeadd04d306f747830e46c0ec577db9a/ClimaMain.f}{\texttt{ClimaMain.f}} specified a H$_2$O $k$-coefficient file that assumed pressure broadening by H$_2$O itself, which is not appropriate for a $N_2$-dominated atmosphere.
Further, the model version also did not allow for convection above \SI{40}{\kilo\meter}.
Consequently, the simulated PT profiles tend to underpredict stratospheric and mesospheric temperatures compared to the expectation from a real Earth twin. 
Newer versions of the \texttt{Atmos} code are not affected by these issues and use updated H$_2$O and CO$_2$  $k$-coefficients \citep{Teal_2022,Vidaurri_2022}.
While this may limit the scientific usefulness of the PT profiles, it should not affect our ability to use them to demonstrate the applicability of our method as such, that is, to show how we can learn efficient, low-dimensional parameterizations of PT profiles from simulated data.

\subsection{\goyal}
This dataset was published by \citet{Goyal_2020} and consists of \num{11293} PT profiles, corresponding to 89 hot Jupiters, each with four recirculation factors, six metallicities and six C/O ratios.%
\footnote{
    The authors report that around \SI{12}{\percent} of the models did not converge; hence the dataset is smaller than the expected \num{12816} combinations.
}
Due to missing data, we manually removed three profiles.
The data were simulated with \texttt{Atmo} \citep{Amundsen_2014,Tremblin_2015, Tremblin_2016,Drummond_2016},%
\footnote{
    We point out that \texttt{Atmos} and \texttt{Atmo} are not the same code.
}
a radiative-convective equilibrium model for atmospheres, using $N = 51$ atmospheric layers.
In addition to the PT profiles, the dataset also includes transmission and emission spectra for each atmosphere, as well as over 250 abundance profiles of different chemical species.

\section{Experiments and results}
\label{sec:experiments}

In this section, we describe the experiments that we have performed to test our proposed approach and show our results.

\subsection{Training and evaluation procedure}
We use stochastic gradient descent to train our models (i.e., one pair of neural networks $E$ and $D$ for each combination of a dataset and a value of $\dim(\z)$) on a random subset of the datasets described in \cref{sec:datasets}, and evaluate their performance on another random subset that is disjoint from the training set. 
For more detailed information about our network architectures, their implementation, and our training procedure, see \cref{sec:implementation-and-training}, or take a look at our code, which is available online.

\subsection{Reconstruction quality}
\label{subsec:reconstruction-quality}

As a first experiment, we study how well our trained model can approximate previously unseen PT profiles from the test set (cf. \cref{par:data-split}), and compare the performance with two baseline methods: (1)~low-order polynomials, and (2)~a PCA-based approach similar to the method of~\citet{Schreier_2020}.

\paragraph{Setup:}
For each $\dim(\z) \in \{1, 2, 3, 4 \}$, we train three instances of our model (using three different random seeds to initialize the neural networks and control the split between training and validation) on the training set.
We do not consider higher values of $\dim(\z)$ here mainly because one of the goals of this work was to demonstrate that our method requires fewer fitting parameters than the baseline methods.
Then, we take the learned decoder and, for each profile in the test set, use nested sampling to find the optimal $\z$ (denoted $\z^*$) to reproduce that profile. 
In this case, \enquote{optimal} means that $\z^*$ minimizes the mean squared error (MSE):
\begin{equation}
    \z^* =
    \argmin_{\z} 
    \frac{1}{N} \cdot \sum_{i=1}^{N} (t_i - D(p_i \given \z))^2 \,.
\end{equation}
We add this additional optimization step to the evaluation to remove the effect of the encoder:
Ultimately, we are only interested in how well $D$ can approximate a given profile for some $\z$---which does not necessarily have to match the output of $E$ for that profile perfectly.
We have chosen nested sampling over gradient-based optimization---which is possible given that $D$ is fully differentiable both with respect to $p$ and $\z$---because the latter is generally not available during an AR, unless one uses a differentiable forward simulator (see, e.g., \texttt{Diff}-$\tau$ from \citealt{Yip_2022}).
Nested sampling, therefore, should give us a more realistic idea of which profiles we can find during a retrieval.
Our optimization procedure is based on \mbox{\emph{UltraNest}}~\citep{Buchner_2021}, with 400 live points and a truncated standard normal prior, limiting each $z_i$ to $\left[ -4, 4 \right]$ (because $\mathcal{L}_\text{norm}$ limits $\lVert \z \rVert$ to $\tau = 3.5$).

For the polynomial baseline, we simply fit each profile in the test set using a polynomial with degree $n_\mathrm{poly} - 1$ for $n_\mathrm{poly} \in \{ 2, 3, 4, 5 \}$.
The minimization objective of the fit is again the MSE.

Finally, for the PCA baseline, we compute a PCA on all the temperature vectors in the training set, and then fit the temperature vectors in the test set as a linear combination of the principal components (PCs), for $n_\mathrm{PC} \in \{ 2, 3, 4, 5 \}$, once again minimizing the MSE.
We note that this PCA baseline must be taken with a grain of salt, for two reasons.
First, it requires us to project all profiles onto a single common pressure grid to ensure that the $i$-th entry of the temperature vectors always has the same interpretation (i.e, temperature at pressure $p_i$).
We do this by linear interpolation.
Consequently, the common pressure grid is determined by the intersection of the pressure ranges of all profiles, meaning that profiles may get clipped both at high and low pressures.
In practice, this affects in particular the \goyal data where, for example, the pressure in the deepest layer varies by more than an order of magnitude between different profiles, whereas for \pyatmos, the profiles all cover a similar pressure range.
Second, combining principal components only returns vectors, not functions, meaning that if we want to evaluate a profile at an arbitrary pressure value, we again have to use interpolation.\looseness=-1

\paragraph{Results:}
We are showing some example profiles and the best-fit reconstruction using our model in \cref{fig:example-profiles-pyatmos}.
The main results of this experiment---the distribution of the (root) mean square error (RMSE) on the test set---are found in \cref{fig:error-distribution-pyatmos}.
The RMSE is simply defined as the square root of the MSE defined in \cref{eq:mse}, that is, the root of the mean squared difference between a given temperature vector and its best-fit reconstruction.
As \cref{fig:error-distribution-pyatmos} shows, our method achieves a significantly lower reconstruction error than both baseline methods, both for the \pyatmos dataset and for the \goyal dataset (see \cref{fig:error-distribution-goyal-2020} in the appendix).
We find that on the \pyatmos dataset, even the one-dimensional version of our model (i.e., each profile is parameterized as just a single number) outperforms the polynomial baseline with five fitting parameters, and is approximately on-par with the PCA solution using five components.

To account for the absolute scale of the data (i.e., Earth-like temperatures vs. hot Jupiters) and allow comparisons across the two datasets, we also compute the mean relative error (MRE):
\begin{equation}
    \text{MRE} = \frac{1}{N} \cdot \sum_{i=1}^{N} \frac{|\hat{t}_i - t_i|}{ t_i } \,.
\end{equation}
We compute this quantity for every profile in the test set, and then take the median of this distribution.
For PCA and our method, the medians are then also mean-averaged over the different runs.
The final results are given in \cref{fig:mre-comparison}.
Looking at the mean relative errors, we note our errors for the \goyal dataset are systematically larger than for the \pyatmos dataset.
We suspect that can be explained as the result of two factors:
First, the \pyatmos training dataset is ten times bigger than the \goyal training dataset, and, second, the \goyal dataset covers a more diverse set of functions compared to the \pyatmos dataset.

\subsection{Interpreting the latent representations of PT profiles in the context of the corresponding atmospheres}
\label{subsec:interpreting-latent-representations}

\begin{figure*}[t]
    \centering
    \includegraphics{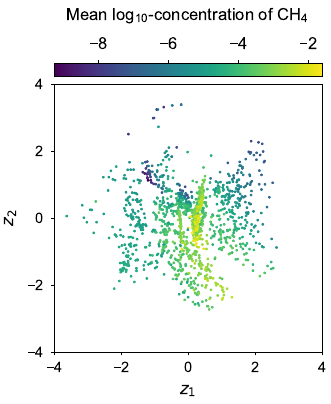}\hfill%
    \includegraphics{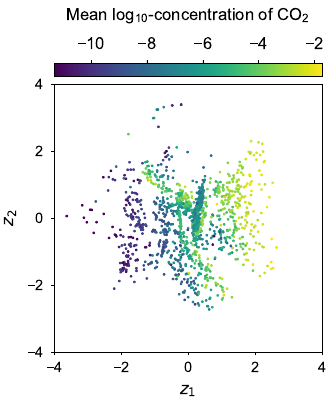}\hfill%
    \includegraphics{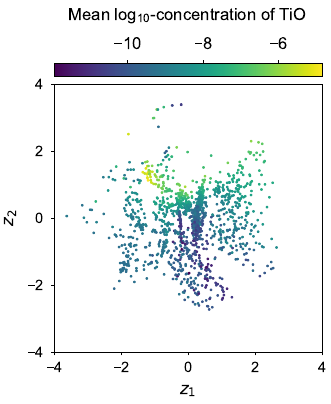}\hfill%
    \caption{
        Examples of 2D latent spaces, where each $\z$ from the test set is color-coded by a property of the corresponding atmosphere.
        Results are shown for the \goyal dataset; additional plots for the \pyatmos are found in \cref{fig:color-coded-z-pyatmos} in the appendix.~%
        \codelink{https://github.com/timothygebhard/ml4ptp/blob/main/scripts/plotting/fig-8-color-coded-z/plot-color-coded-z.py}
    }
    \label{fig:color-coded-z-goyal-2020}
\end{figure*}

In this section, we take a closer look at the latent representations of PT profiles that our models learn and the extent to which they lend themselves to interpretation.
We begin with \cref{fig:latent-space}, where we present an illustration of our entire method using a model with $\dim(\z)=2$ that we trained on the \goyal dataset.
The upper part of the figure shows how the encoder $E$ maps PT profiles---which are given by vectors $\bm{p}$ and $\bm{t}$, which in this example are 51-dimensional---onto 2-dimensional latent representations $\z$.
The three colored dots in the latent space correspond to the three PT profiles shown on top, which are examples from our test set (i.e., profiles that were not used for training).
The gray dots in the latent space show the representations of all PT profiles in our test set.
Their distribution approximately follows a 2D standard Gaussian (cf. the marginal distributions), which is precisely what we would expect, as this was the reason we introduced the $\mathcal{L}_\text{MMD}$ term to the loss function.
The bottom part of the figure illustrates what happens if we place a regular grid on the latent space and, for each $\z$ on the grid, evaluate $D(\cdot \given \z)$ on a fixed pressure vector $\bm{p}'$, which does not have to match the $\bm{p}$ of any profile in the training or test dataset.
For example, $\bm{p}'$ can have a higher resolution (i.e., more atmospheric layers) than the training data.

In both parts of the figure, we observe that PT profiles with different shapes---implying different physical processes---are mapped to different parts of the latent space, and that similar PT profiles are grouped together in the latent space.
Here, \enquote{similar} refers not only to the shape of the PT profiles, but also to the properties of the corresponding atmospheres.
To see this, we turn to \cref{fig:color-coded-z-goyal-2020}.
As mentioned above, both the \pyatmos and \goyal datasets contain not only PT profiles, but also various other properties of the corresponding atmospheres, such as concentrations or fluxes of chemical species.
In \cref{fig:color-coded-z-goyal-2020}, we show several scatter plots of the latent representations of the test set (i.e., the $\z$ we obtained in the first experiment), color-coded according to some of these atmospheric properties---for example, the mean concentration of CO\textsubscript{2} in the atmosphere corresponding to a given PT profile.
For ease of visualization, we limit ourselves to the case $\dim(\z) = 2$.
We observe that there are clear patterns in these scatter plots that show a strong correlation between our latent representations and the properties of the corresponding atmospheres, again demonstrating that PT profiles from physically similar atmospheres are grouped together.

With this in mind, we return to the lower half of \cref{fig:latent-space}, where we evaluate $D$ on a grid of $\z$ values.
We observe that not only do different shapes of PT profiles correspond to different parts of the latent space, but the PT profiles given $D(\cdot \given \z)$ also vary smoothly as a function of $\z$.
This means that the decoder not only reproduces the real PT profiles that were used to train it, but also allows smooth interpolation between them.
Along the axes of the latent space, one can identify specific behaviors: 
For example, fixing $z_2 = 0$ and increasing $z_1$ leads to cooling in the upper atmosphere, while fixing $z_1 = 2$ and increasing $z_2$ leads to heating that extends further into the upper layers. 
Profiles around the center of the latent space (i.e., $z_1 \approx z_2 \approx 0$) are almost isothermal, indicating a relatively uniform temperature distribution throughout the atmosphere.
A negative value of $z_1$ is generally associated with the presence of a thermal inversion, and for these cases, $z_2$ seems to encode the altitude where the inversion happens. 
For instance, for $z_1 = -2$ the profile shows an inversion in the mid-atmosphere (around $\log_{10}(P) = -2$) for $z_2 = 2$, whereas for $z_2 = -2$, the inversion only happens at very high altitudes/low pressures around $\log_{10}(P) = -6$. 

Finally, we can also draw direct connections between the behavior of the PT profiles in latent space and the correlations of $\z$ with some of the atmospheric parameters that we show in \cref{fig:color-coded-z-goyal-2020}.
For example, a high concentration of \ch{TiO} (panel 3 of \cref{fig:color-coded-z-goyal-2020}) corresponds to the hot, inverted atmospheres in the upper left of the latent space.
This is consistent with expectations from the literature, where \ch{TiO} is indeed one of the canonical species proposed to cause temperature inversions in hot Jupiters \citep[e.g.,][]{Hubeny_2003, Fortney_2008, Piette_2020}.

Overall, this suggests that the latent space is to some extent interpretable:
While the latent variables may not correspond directly to known physical parameters---as is the case, for example, for the analytical parameterizations of \citet{Madhusudhan_2009} and \citet{Guillot_2010}---their relationships with other variables may provide insight into the behavior of exoplanet atmospheres.

Finally, a small word of caution: While we have just shown that the latent space is \emph{in principle} interpretable, any \emph{specific} interpretation---that is, a statement along the lines of \enquote{PT profiles with hot upper layers are found in the top left corner of the latent space}---will only hold for one specific model.
If we re-train that same model using a different random seed to initialize the neural network, or using a different dataset, we will end up with an equivalent but different latent space.
This is because our model is, in general, not identifiable, and there exists no preferred system of coordinates for $\z$: For example, we can rotate or mirror the latent space without losing information, and unless we add additional constraints to the objective function, there is no reason why the model should favor any particular orientation.
However, this simply means that the value of $\z$ can only be interpreted for a given trained model. 
It does not, in any way, affect the ability of our model to be used for atmospheric retrievals.\looseness=-1

\subsection{Usage in an atmospheric retrieval}
\label{subsec:retrieval-experiment}

A major motivation for the development of our method is the possibility of using it for atmospheric retrievals.
In this experiment, we show how we can easily plug our trained model into an existing atmospheric retrieval pipeline.
The target PT profile we use for this experiment is taken from the \pyatmos test set.
We made sure that it is covered by the training data (i.e., there are similar PT profiles in the training set); however, in order not to make it too easy for our method, we chose a PT profile shape that is comparatively rare in the training data (see below).
In \cref{sec:additional-retrievals}, we consider an even more challenging setting and show two additional results for PT profiles that were generated with a different code and that are not covered by our training distribution.

\paragraph{Setup:}
For our demonstration, we add our trained model to the atmospheric retrieval routine first introduced in \citet{Konrad_2022} and developed further in \citet{Alei_2022b} and \citet{Konrad_2023}, which is based on \texttt{petitRADTRANS} \citep{Molliere_2019}, \texttt{LIFEsim} \citep{Dannert_2022} and \texttt{PyMultiNest} \citep{Buchner_2014}. 
The goal of this experiment is to compare the PT profile parameterization capabilities of a version of our model (trained on the \pyatmos dataset using $\dim(\z) = 2$) with the fourth-order polynomial (i.e., five fitting parameters) used in \citet{Konrad_2022} and \citet{Alei_2022b}.
We perform two cloud-free atmospheric retrievals (using both our PT model or the polynomial one) on the simulated thermal emission spectrum of an Earth-like planet orbiting a Sun-like star at a distance of \SI{10}{\parsec}, assuming a photon noise signal-to-noise ratio of 10 at a wavelength of $\lambda = \SI{11. 2}{\micro\meter}$.
The wavelength range was chosen as $[\SI{3}{\micro\meter}, \SI{20}{\micro\meter}]$, and the spectral resolution was $R = \nicefrac{\lambda}{\Delta \lambda} = 200$. 
The number of live points for \texttt{PyMultiNest} was set to \num{700}.
In addition to the parameters required for the PT profile, we retrieve 10 parameters of interest; see \cref{tab:retrieval-parameters} for an overview.
This setup closely matches \citet{Konrad_2022} and \citet{Alei_2022b}.

For the retrieval with our model, we use a 2-dimensional uniform prior, $\mathcal{U}_2(-4, 4)$ for $\z$.
The reason for choosing this uniform prior instead of a Gaussian is that the shape of the ground truth PT profile is relatively under-represented in the \pyatmos dataset that we use for training.
This means that our trained model, together with a Gaussian prior on $\z$, assigns it a very small prior probability, making it difficult for the retrieval routine to find the correct PT profile.
Using a uniform prior is an easy way to give more weight to \enquote{rare} profiles.
We discuss this choice, as well as potential alternatives, in more detail \cref{subsec:choosing-a-prior-for-z}.

\begin{table*}
\centering
\caption{
    Overview of all parameters in our simulated atmospheric retrieval, including the respective priors, true values, and retrieved values for both PT profile parameterization schemes.
    For the latter, we report the median as well as the 2.5\% and 97.5\% percentiles.
}
\label{tab:retrieval-parameters}
\footnotesize
\renewcommand{\arraystretch}{1.3}
\newcommand{\range}[2]{\raisebox{0.6mm}{\ensuremath{_{-#1}^{+#2}}}}
\begin{tabular}{lll S[table-format=3.3] S[table-format=7.2]@{\hspace{1mm}}>{\tiny}l S[table-format=7.2]@{\hspace{1mm}}>{\tiny}l}
\toprule
Parameter                   &  Description                  &  Prior                         &  {True value}  &  \multicolumn{2}{c}{Retrieved (polynomial)} & \multicolumn{2}{c}{Retrieved (our method)} \\
\midrule
$a_4$                       &  PT profile parameter (polynomial)  &  $\mathrm{Uniform}(0, 10)$     &         {---}  &       1.33 &     $\range{0.67}{0.39}$  &       \multicolumn{2}{c}{---} \\
$a_3$                       &  PT profile parameter (polynomial)  &  $\mathrm{Uniform}(0, 100)$    &         {---}  &      30.49 &    $\range{27.32}{49.83}$ &       \multicolumn{2}{c}{---} \\
$a_2$                       &  PT profile parameter (polynomial)  &  $\mathrm{Uniform}(0, 300)$    &         {---}  &     106.78 &  $\range{87.47}{159.09}$  &       \multicolumn{2}{c}{---} \\
$a_1$                       &  PT profile parameter (polynomial)  &  $\mathrm{Uniform}(0, 500)$    &         {---}  &     204.71 & $\range{118.24}{239.09}$  &       \multicolumn{2}{c}{---} \\
$a_0$                       &  PT profile parameter (polynomial)  &  $\mathrm{Uniform}(0, 600)$    &         {---}  &     384.86 &  $\range{90.09}{178.40}$  &       \multicolumn{2}{c}{---} \\
$z_1$                       &  PT profile parameter (our method)  &  $\mathrm{Uniform}(-4, 4)$     &         {---}  &               \multicolumn{2}{c}{---}  &  -2.56 & $\range{1.34}{1.69}$ \\
$z_2$                       &  PT profile parameter (our method)  &  $\mathrm{Uniform}(-4, 4)$     &         {---}  &               \multicolumn{2}{c}{---}  &  -0.04 & $\range{1.95}{1.19}$ \\
$\log_{10}(P_0)$            &  Surface pressure (in bar)          &  $\mathrm{Uniform}(-2, 2)$     &         0.013  &      -0.48 &     $\range{0.61}{0.60}$  &  -0.05 & $\range{0.16}{0.14}$ \\
$R_\mathrm{pl}$             &  Planet radius (in $R_\oplus$)      &  $\mathrm{Normal}(1, 0.2)$     &         1.00   &       1.00 &     $\range{0.03}{0.03}$  &   1.00 & $\range{0.03}{0.03}$ \\
$\log_{10}(M_\mathrm{pl})$  &  Planet mass (in $M_\oplus$)        &  $\mathrm{Normal}(0, 0.4)$     &         0.00   &       0.02 &     $\range{0.59}{0.57}$  &   0.03 & $\range{0.64}{0.65}$ \\
$\log_{10}(X_{\ch{N_2}})$   &  \ch{N_2} mass fraction             &  $\mathrm{Uniform}(-2, 0)$     &        -0.11   &      -1.04 &     $\range{0.89}{0.93}$  &  -1.05 & $\range{0.88}{0.93}$ \\
$\log_{10}(X_{\ch{O_2}})$   &  \ch{O_2} mass fraction             &  $\mathrm{Uniform}(-2, 0)$     &        -0.70   &      -1.10 &     $\range{0.85}{0.96}$  &  -1.11 & $\range{0.84}{0.99}$ \\
$\log_{10}(X_{\ch{CO_2}})$  &  \ch{CO_2} mass fraction            &  $\mathrm{Uniform}(-10, 0)$    &        -3.40   &      -2.72 &     $\range{1.25}{1.31}$  &  -3.22 & $\range{0.75}{0.80}$ \\
$\log_{10}(X_{\ch{CH_4}})$  &  \ch{CH_4} mass fraction            &  $\mathrm{Uniform}(-10, 0)$    &        -5.77   &      -5.06 &     $\range{1.27}{1.26}$  &  -5.61 & $\range{0.84}{0.84}$ \\
$\log_{10}(X_{\ch{H_2O}})$  &  \ch{H_2O} mass fraction            &  $\mathrm{Uniform}(-10, 0)$    &        -3.00   &      -2.29 &     $\range{1.32}{1.36}$  &  -2.82 & $\range{0.73}{0.77}$ \\
$\log_{10}(X_{\ch{O_3}})$   &  \ch{O_3} mass fraction             &  $\mathrm{Uniform}(-10, 0)$    &        -6.52   &      -6.00 &     $\range{1.04}{1.07}$  &  -6.39 & $\range{0.69}{0.72}$ \\
$\log_{10}(X_{\ch{CO}})$    &  \ch{CO} mass fraction              &  $\mathrm{Uniform}(-10, 0)$    &        -6.90   &      -7.47 &     $\range{2.40}{2.93}$  &  -7.76 & $\range{2.13}{2.45}$ \\
\bottomrule
\end{tabular}
\end{table*}

\begin{figure*}
    \placeonpage{11}
    \centering
    \begin{subfigure}[t]{90mm}
        \centering
        \vphantom{69.9mm}%
        \includegraphics[width=90mm]{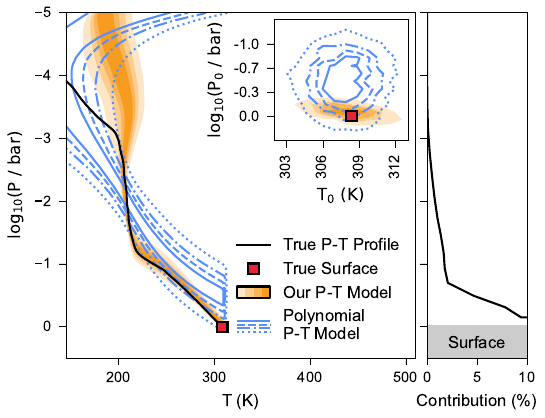}
        \label{fig:retrieved-pt-profile}
    \end{subfigure}%
    \hfill%
    \begin{subfigure}[t]{90mm}
        \centering
        \vphantom{69.9mm}%
        \includegraphics[width=90mm]{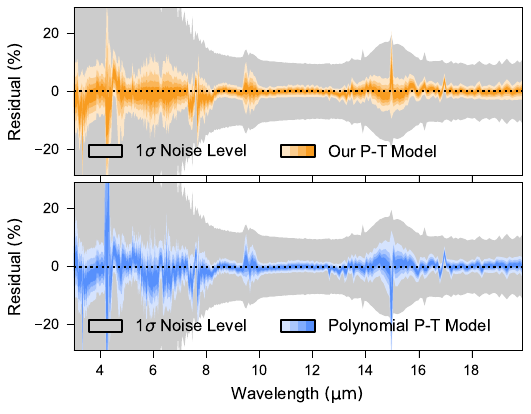}
        \label{fig:retrieved-spectrum}
    \end{subfigure}%
    \caption{
        Results for our simulated atmospheric retrieval of an Earth-like planet using \texttt{petitRADTRANS}, \texttt{LIFEsim}, and \texttt{PyMultiNest}. 
        \emph{Left:} The retrieved PT profiles for the polynomial baseline and our model, together with the emission contribution function.
        \emph{Right:} The relative fitting error of the spectrum, computed as (true spectrum $-$ simulated spectrum) / true spectrum.
    }
    \label{fig:retrieval-results}
\end{figure*}

\paragraph{Results:}
We show the main results of this experiment in the form of the retrieved PT profile and spectrum residuals in \cref{fig:retrieval-results}.
First, we find that the spectrum residuals demonstrate that both retrievals reproduce the simulated spectrum with sufficient accuracy, despite the differences in the PT parametrization. 
The residual's quantiles are centered on the truth, and are significantly smaller than the photon noise-level. 
Second, for the retrieved PT profiles, we visually find that the result obtained with our model---unlike the polynomial baseline---is in good agreement with the ground truth, except at the highest layers of the atmosphere, where it tends to underestimate the true temperature.
We believe this can be explained by the fact that the upper layers of the atmosphere (i.e., low pressures) have little to no effect on the simulated exoplanet spectrum (cf. the emission contribution functions in the right panel of \cref{fig:retrieved-pt-profile}), making it hard for the retrieval routine to constrain the PT structure in the upper atmosphere.
Third, the retrieved constraints for the surface pressure $P_0$ and temperature $T_0$, are much tighter and more accurate for our model than the polynomial baseline.
Of course, this is also partly because our model represents a much narrower prior over the PT profiles that we are willing to accept as an explanation of the data (i.e., the spectrum) compared to the polynomial.
This is an assumption, and in this case, it is justified by construction.
In general, however, the decision of what is an appropriate prior for a given atmospheric retrieval problem is beyond the scope of the method as such.
We discuss this in more detail in \cref{subsec:role-of-training-dataset}.
Finally, we note that the retrieval using our model was significantly faster than the baseline: by decreasing the number of parameters required to fit the PT profile from five (polynomial) to two (our model), we reduced the total runtime by a factor of \num{3.2}.

\section{Discussion}

In this section, we discuss the advantages and limitations of our method and suggest directions for future work.

\subsection{Advantages}
Our approach for parameterizing PT profiles has multiple potential benefits compared to existing approaches.
First, our model makes realistic, physically consistent PT profiles---which require a computationally expensive solution of radiative transfer equations---available as an ad hoc fitting function.
It works, in principle, for all types of planets (Earth-like, gas giants, etc.) as long as we can create a suitable training dataset through simulations. 
In this context, \enquote{suitable} refers to, for example, the total size of the dataset and the density in parameter space (cf. the single out-of-distribution profile for \pyatmos). 
Our method then basically provides a tool which allows using the distribution of the PT profiles in that training set as a prior for the PT profile during a retrieval.
Second, during an atmospheric retrieval, the parameterization of our model uses a simple prior for $\z$ that does not require hard-to-interpret fine-tuning (see also \cref{subsec:choosing-a-prior-for-z}).
Third, as we have shown in \cref{sec:experiments}, our method can fit realistic PT profiles with fewer parameters than the baseline and still achieve a lower fit error.
Limiting the number of parameters needed to express a PT profile during retrieval can lead to faster processing times, or conversely, allow the retrieval of additional parameters when operating within a fixed computational budget.
We would like to emphasize at this point that accelerating atmospheric retrievals is not only relevant for the analysis of existing observational data (e.g., from the James Webb Space Telescope), but can also be beneficial during the planning stages of future exoplanet science missions, such as LIFE~\citep{Quanz_2022} or the recently announced Habitable Worlds Observatory \citep{Clery_2023}, which require the simulation of thousands of retrievals.

Finally, looking to the future, it seems plausible that approaches that replace the entire Bayesian inference routine with machine learning may gain further traction in the context of atmospheric retrievals (see, e.g., \citealt{MarquezNeila_2018}, \citealt{Soboczenski_2018}, \citealt{Cobb_2019}, \citealt{ArdevolMartinez_2022}, \citealt{Yip_2022}, or \citealt{Vasist_2023}).
In this case, reducing the number of retrieval parameters may no longer be a major concern--once trained, the inference time of these models is relatively independent of the number of parameters.
However, even in this case, our approach will still be useful to provide a parameterized description of the posterior over physically consistent PT profiles.

\subsection{The role of the training dataset}
\label{subsec:role-of-training-dataset}
The practical value of our method stands and falls with the dataset that is used to train it.
This should come as no surprise, but there are a few aspects to this that we will discuss in more detail here.

First, it is important to understand that the training dataset defines a distribution over the functions that our model learns to provide parameterized access to.
When using our model in an atmospheric retrieval, this distribution then becomes the prior over the set of PT profiles to be considered.
Compared to, for example, low-order polynomials, this is a much narrower prior: While polynomials can fit virtually anything (this is the idea of a Taylor series), our model will only produce outputs that resemble the PT profiles that it encountered during training.
If this is a good fit for a given target profile, we can expect to significantly outperform the broader prior of polynomials. 
Conversely, if our prior does not cover the target profile, we cannot expect good results.
However, this is not a limitation of our method in particular, but is true for all approaches that use a parameterization for the PT profile instead of, for example, a level-by-level radiative transfer approach.
As \citet{Line_2013} explain, using \enquote{[a] parameterization does force the retrieved atmospheric temperature structure to conform only to the profile shapes and physical approximations allowed by that parameterization.}
Thus, as always in Bayesian inference, the decision to use our method as a prior for the PT profile (rather than, e.g., low-order polynomials) represents an assumption about the space of possible explanations for the data (e.g., the observed spectrum) that one is willing to accept, and it is up to the scientist performing an atmospheric retrieval to decide whether that assumption is justified.

Finally, it is important to keep in mind that our model will learn to replicate any anomalies and imbalances present in the training data. 
For example, if atmospheres without inversions are overrepresented in the training data (compared to atmospheres with inversions), our learned function distribution will also place a higher probability on PT profiles without inversions, which will obviously affect the results of a retrieval.

\subsection{Choosing a prior distribution for \textbf{z}}
\label{subsec:choosing-a-prior-for-z}

The last point in the previous section is closely related to the question of which prior one should choose for $\z$ during a retrieval.
The natural choice, of course, is to assume a $d$-dimensional standard Gaussian prior for $\z$ as that is the distribution that we encourage during training by using the $\mathcal{L}_\text{MMD}$ loss.
However, there are a few additional considerations to keep in mind here.
First, using a Gaussian prior for $\z$ means that the induced distribution over PT profiles, given by $D(\cdot \given \z)$, will (approximately) match the distribution in the training dataset.
This is a good choice if the training dataset itself constitutes a good prior for the PT profiles one wants to consider during a retrieval.
On the other hand, consider now the case mentioned above where the training dataset contains all the expected shapes, but is highly biased towards one of the modes, not because this reflects the (assumed) true distribution of atmospheres, but as an artifact of the dataset generation process.
In this case, it may make sense to choose a non-Gaussian prior (e.g., a uniform distribution) for $\z$ to compensate for this imbalance.
When doing so, it is important to ensure that $\z$ stays in the value range that the decoder $D$ encountered during training (i.e., $\lVert \z \rVert < \tau$).
Another (and perhaps more principled) way to solve this problem is to resample the training dataset to balance the different modes (e.g., \enquote{inversion} and \enquote{no inversion}).
For our experiment in \cref{subsec:retrieval-experiment}, we have validated that such a resampling-based approach does indeed work, but have chosen not to include all the details of this additional experiment so as not to distract too much from the main message of our paper.

Finally, there is the fact that our trained model does not only return physically consistent PT profiles, but also allows smooth interpolation (or, to some extent, extrapolation) between them.
In cases where this is not desired, for example because one wants to be sure that only PT profiles very close to the training dataset are produced, one can choose a data-driven prior distribution for \z. 
A simple way to achieve this is to train a normalizing flow (see, e.g., \citealt{Kobyzev_2021} or \citealt{Papamakarios_2021} for recent introductions) to match the distribution of $\z$'s obtained from the training dataset.
If we then use the normalizing flow to sample \z, we will almost perfectly reproduce the distribution of the training data set.
However, this can also exacerbate some of the problems we discussed earlier, for example, if the training data is unbalanced.

\subsection{Directions for future research}
\label{subsec:future-directions}

We see several directions for future work.
For example, to account for the fact that not all parts of the atmosphere contribute equally to the observed spectrum, one possible approach could be to use a modified reconstruction loss weighted by the respective contribution function for each point in the profile.
In the case of emission spectra, this could focus the expressiveness of the model on regions of higher pressure, which have a greater influence on the spectrum, rather than regions at high altitudes, where the thermal structure of the atmosphere is difficult to constrain from the available observational data.
For transmission spectra, which probe the upper layers of the atmosphere, one would do the opposite and give more weight to high altitudes.

Another promising direction could be to extend our method to parameterize not only temperature as a function of pressure, but also chemical abundance profiles (e.g., the concentration of \ch{CO_2} in each layer), to make them accessible during a retrieval.

Finally, given a sufficiently large grid of training data, our approach can also be used to parameterize the thermal structure of 3-dimensional atmospheres, where temperature depends not only on pressure but also on the longitude and latitude.

\section{Summary and conclusion}

In this work, we have introduced a new approach to parameterizing PT profiles in the context of atmospheric retrievals of exoplanets.
Unlike existing approaches for this problem, we do not make any a priori assumptions about the family of functions that describe the relation between $P$ and $T$, but instead learn a distribution over such functions from data.
In our framework, PT profiles are represented by low-dimensional real vectors that can be used to condition a particular neural network, which, once conditioned, \emph{is} the corresponding PT profile (i.e., a single-argument, single-output function that maps pressure onto temperature values).

We experimentally demonstrated that our proposed method works for both terrestrial planets and hot Jupiters, and when compared with existing methods for parameterizing PT profiles, we found that our methods give better approximations on average (i.e., smaller fitting errors) while requiring a smaller number of parameters.
Furthermore, we showed that our learned latent spaces are still interpretable to some extent: For example, the latent representations are often directly correlated with other properties of the atmospheres, meaning that physically similar profiles are grouped together in the latent space.
Finally, we have demonstrated by example that our method can be easily integrated into an existing atmospheric retrieval framework, where it resulted in a significant speedup in terms of inference time, and produced a tighter and more accurate posterior for the PT profile than the baseline.

Given access to a sufficient training dataset, we believe that our method has the potential to become a valuable tool for exoplanet characterization: Not only could it allow the use of physically consistent PT profiles during a retrieval, but it can also reduce the number of parameters required to model the thermal structure of an atmosphere. 
This can either reduce the overall computational cost or free up resources for other retrieval parameters of interest, allowing for more efficient and accurate atmospheric retrievals, which in turn could improve our understanding of exoplanetary habitability and the potential for extraterrestrial life.

\begin{notes}{Code and data availability}
    All code for our method and experiments is available here: \url{https://github.com/timothygebhard/ml4ptp}.
    Our datasets and the final trained models are available here: \url{https://doi.org/10.17617/3.K2CY3M}.
\end{notes}

\begin{notes}{Used software}
    This work has made use of a number of open-source Python packages, including
    \texttt{matplotlib} \citep{Hunter_2007},
    \texttt{numpy} \citep{Harris_2020},
    \texttt{scipy} \citep{Virtanen_2020},
    \texttt{scikit-learn} \citep{Pedregosa_2011},
    \texttt{pymultinest} \citep{Buchner_2014},
    \texttt{pandas} \citep{McKinney_2010},
    \texttt{petitRADTRANS} \citep{Molliere_2019},
    \texttt{pytorch} \citep{Paszke_2019}, and
    \texttt{ultranest} \citep{Buchner_2021}.
    The colorblind-friendly color schemes in our figures are based on \citet{Petroff_2021}.
\end{notes}

\begin{acknowledgements}
    We thank Sandra T. Bastelberger and Nicholas Wogan for pointing out the potential problems with the PyATMOS simulations and taking the time to discuss the issue with us.
    Furthermore, we thank Markus Bonse, Felix Dannert, Maximilian Dax, Emily Garvin, Jean Hayoz and Vincent Stimper for their helpful comments on the manuscript. 
    
    Part of this work has been carried out within the framework of the National Centre of Competence in Research PlanetS supported by the Swiss National Science Foundation under grants \mbox{51NF40\_182901} and \mbox{51NF40\_205606}. 
    SPQ and EA acknowledge the financial support of the SNSF.
    TDG acknowledges funding through the Max Planck ETH Center for Learning Systems.
    BSK acknowledges funding through the European Space Agency’s Open Space Innovation Platform (ESA OSIP) program.
\end{acknowledgements}

    \bibliographystyle{aa}
    \bibliography{main.bib}
    \flushcolsend

    \clearpage
    \appendix

\section{Method}
\label{app:method}

In this appendix, we provide additional background information for \cref{sec:method} (\enquote{Method}) of the main part of the paper.

\subsection{Loss functions}
To encourage the encoder to produce values of $\z$ that follow a given prior distribution (in our case: a $d$-dimensional Gaussian), we have chosen to add a term to the loss based on the Maximum Mean Discrepancy (MMD): $\mathcal{L}_\text{MMD}$. 
Our motivation for this choice came from \citet{Zhao_2017}, who have shown that for the case of variational auto-encoders (VAE)---a method that is conceptually related to our approach---the usage of the MMD maximizes the mutual information between the latent representation~$\z$ and the output of the decoder~$D$.
Given our main task (i.e., atmospheric retrievals with simulation-based Bayesian inference), this seems desirable: 
After all, we do not only want to generate random PT profiles from the distribution given by the training data, but we would like to explore the latent space systematically (e.g., using nested sampling) to find the pressure-temperature profile that is the best match for our data.

There are, of course, also alternatives to the MMD which one could explore in future work.
For example, one might choose to minimize the Kullback-Leibler (KL) divergence between $\z$ and a standard multivariate Gaussian.
This corresponds to the loss function used in the original paper on neural processes~\citep{Garnelo_2018b}, who use the evidence lower bound (ELBO) as the objective function to train their models.
However, at least in the case of variational auto-encoders, it has been shown (see, e.g., \citealt{Chen_2016} and \citealt{Zhao_2017}) that for sufficiently powerful models, using the ELBO objective can lead to a decoder that actually ignores the latent variable; a behavior that does not seem helpful for the application that we have in mind.

\subsection{Maximum Mean Discrepancy (MMD)}
\label{subsec:mmd}

The MMD is a kernel-based approach for estimating the distance between two probability distributions.
It can be computed from two samples and works in arbitrary dimensions.
Following \citet{Borgwardt_2006} and \citet{Gretton_2012}, an unbiased estimator of the MMD of two samples $X = \lbrace x_1, \ldots, x_m \rbrace$ and $Y = \lbrace y_1, \ldots, y_n \rbrace$, with all $x_i$ and $y_i$ coming from the same vector space $\mathcal{X}$, is given by:
\begin{equation*}
\begin{split}
    \mathrm{MMD}^2(X, Y) ={} 
    &\frac{1}{m \cdot (m - 1)} \cdot \sum_{i=1}^{m} \sum_{j \neq i}^{m} k(x_i, x_j) \,+ \\
    &\frac{1}{n \cdot (n - 1)} \cdot \sum_{i=1}^{n} \sum_{j \neq i}^{n} k(y_i, y_j) \,- \\
    &\frac{2}{m \cdot n} \cdot \sum_{i=1}^{m} \sum_{j = 1}^{n} k(x_i, y_j) \,.
\end{split}
\end{equation*}
Here, $k(x, y): \mathcal{X} \times \mathcal{X} \to \R$ is a kernel function.
In our application, we use a Gaussian kernel:
\begin{equation*}
    k(x, y) = \exp \left\{ -\frac{\lVert x - y \rVert^2_2}{2 \sigma^2} \right\} \,,
\end{equation*}
where we estimate $\sigma$ using the median heuristic from \citet{Gretton_2012}, that is, we set $\sigma$ to the median of the distances between all points in the aggregate sample $X \cup Y$.
We note that our loss term $\mathcal{L}_\text{MMD}$ actually uses the squared MMD, as given by the above formula.

\section{Implementation and training}
\label{sec:implementation-and-training}

In this appendix, we provide more detailed information about the neural networks that we use for the encoder and decoder, their technical implementation, as well as our training procedure.

\subsection{Network architectures}
\begin{figure*}
    \begin{subfigure}[b]{0.5\linewidth}
        \centering
        \begin{tikzpicture}

    \tikzset{every node/.append style={node font=\sffamily}}
    \tikzstyle{arrow}=[->, shorten >= 1mm, shorten <= 1mm]

    \matrix (pt) [left delimiter={[},right delimiter={]}, matrix of math nodes, inner sep=1mm] 
         { \textbf{p} & \\ \textbf{t} & \\};

    %\node [draw, minimum width=60mm, minimum height=8mm, thick] (pt) {};
    %\node [align=center, fill=white, minimum width=58mm,] at (pt) {\textbf{p} \\ \textbf{t}};
    %\draw (pt.center)+(0.25, 0.225) -- ++(2.75, 0.225);
    %\draw (pt.center)+(0.25, -0.225) -- ++(2.75, -0.225);
    %\draw (pt.center)+(-0.25, 0.225) -- ++(-2.75, 0.225);
    %\draw (pt.center)+(-0.25, -0.225) -- ++(-2.75, -0.225);

    \node [left=16mm of pt, anchor=center, font=\tiny, align=center, gray] {Input PT profile:\\ pressures and\\ temperatures\\ as 2 channels};

    \node [below=3.5mm of pt, draw, dotted, minimum width=64mm, minimum height=34mm, rounded corners=2.5mm] (convnet) {};
    \node [left=2.5mm of convnet, anchor=center, rotate=90, font=\tiny, gray] {Convnet};
    
    \node [below=6mm of pt, draw, minimum width=60mm, font=\tiny] (c1) {Conv1D layer (2 $\to$ 64) + LeakyReLU};
    \node [below=3mm of c1, draw, minimum width=60mm, font=\tiny] (c2) {Conv1D layer (64 $\to$ 64) + LeakyReLU};
    \node [below=3mm of c2, draw, minimum width=60mm, font=\tiny] (c3) {Conv1D layer (64 $\to$ 64) + LeakyReLU};
    \node [below=3mm of c3, draw, minimum width=60mm, font=\tiny] (c4) {Conv1D layer (64 $\to$ 1) + LeakyReLU};

    \node [below=5.5mm of c4, draw, dotted, minimum width=64mm, minimum height=42mm, rounded corners=2.5mm] (mlp) {};
    \node [left=2.5mm of mlp, anchor=center, rotate=90, font=\tiny, gray] {MLP};
    
    \node [below=8mm of c4, draw, minimum width=60mm, font=\tiny] (l1) {Linear layer (N $\to$ 512) + LeakyReLU};
    \node [below=3mm of l1, draw, minimum width=60mm, font=\tiny] (l2) {Linear layer (512 $\to$ 512) + LeakyReLU};
    \node [below=3mm of l2, draw, minimum width=60mm, font=\tiny] (l3) {Linear layer (512 $\to$ 512) + LeakyReLU};
    \node [below=3mm of l3, draw, minimum width=60mm, font=\tiny] (l4) {Linear layer (512 $\to$ 512) + LeakyReLU};
    \node [below=3mm of l4, draw, minimum width=60mm, font=\tiny] (l5) {Linear layer (512 $\to$ d)};

    \node [below=6mm of l5] (z) {\textbf{z}};
    
    \draw [arrow] (pt) -- (c1);
    \draw [arrow] (c1) -- (c2);
    \draw [arrow] (c2) -- (c3);
    \draw [arrow] (c3) -- (c4);
    \draw [arrow] (c4) -- (l1);
    \draw [arrow] (l1) -- (l2);
    \draw [arrow] (l2) -- (l3);
    \draw [arrow] (l3) -- (l4);
    \draw [arrow] (l4) -- (l5);
    \draw [arrow] (l5) -- (z);

\end{tikzpicture}
        \caption{Encoder architecture.}
    \end{subfigure}%
    \begin{subfigure}[b]{0.5\linewidth}
        \centering
        \begin{tikzpicture}

    \makeatletter
    \DeclareRobustCommand{\rvdots}{%
      \vbox{
        \baselineskip4\p@\lineskiplimit\z@
        \kern-\p@
        \hbox{.}\hbox{.}\hbox{.}
      }}
    \makeatother

    \tikzset{every node/.append style={node font=\sffamily}}
    \tikzstyle{arrow}=[->, shorten >= 1mm, shorten <= 1mm]

    \node (p) {p\textsubscript{i}\vphantom{lp}};
    \node [above=0mm of p, gray, font=\tiny, align=center] {Input\\pressure};
    
    \node [left=25mm of p] (z) {\textbf{z}\vphantom{lp}};
    \node [above=0mm of z, gray, font=\tiny, align=center] {Latent\\representation};

    \node [below=5mm of p, circle, draw, inner sep=0.5mm] (concat1) {+};
    \node [right=1mm of concat1, gray, font=\tiny] {concatenate};

    \node [below=4mm of concat1, draw, minimum width=54mm, font=\tiny] (layer1) {Linear layer 1 + LeakyReLU};
    \node [below=4mm of layer1, circle, draw, inner sep=0.5mm] (concat2) {+};
    \node [right=1mm of concat2, gray, font=\tiny] {concatenate};
    
    \node [below=4mm of concat2, draw, minimum width=54mm, font=\tiny,] (layer2) {Linear layer 2 + LeakyReLU};
    \node [below=4mm of layer2] (dots) {$\rvdots$};
    \node [below=4mm of dots, circle, draw, inner sep=0.5mm] (concatn) {+};
    \node [right=1mm of concatn, gray, font=\tiny] {concatenate};
    
    \node [below=4mm of concatn, draw, minimum width=54mm, font=\tiny,] (layern) {Linear layer 5};
    \node [below=5mm of layern] (t) {$\hat{\textsf{t}}$\textsubscript{i}};
    \node [below=0mm of t, gray, font=\tiny] {Predicted temperature};
    
    \draw [arrow] (p) -- (concat1); 
    \draw [arrow] (z) |- node [pos=0.75, rotate=0, font=\tiny, fill=white, draw, inner sep=0.5mm] {M\textsubscript{1}} (concat1);
    \draw [arrow] (z) |- node [pos=0.75, rotate=0, font=\tiny, fill=white, draw, inner sep=0.5mm] {M\textsubscript{2}} (concat2);
    \draw [arrow] (z) |- node [pos=0.75, rotate=0, font=\tiny, fill=white, draw, inner sep=0.5mm] {M\textsubscript{5}} (concatn);
    \draw [arrow] (concat1) -- (layer1);
    \draw [arrow] (layer1) -- (concat2);
    \draw [arrow] (concat2) -- (layer2);
    \draw [arrow] (layer2) -- (dots);
    \draw [arrow] (dots) -- (concatn);
    \draw [arrow] (concatn) -- (layern);
    \draw [arrow] (layern) -- (t); 

\end{tikzpicture}
        \caption{Decoder architecture.}
    \end{subfigure}%
    \caption{
        Schematic illustration of the architecture of our encoder and decoder networks.
        The numbers in parentheses denote the number of input and output features (or channels) of a layer.
    }
    \label{fig:encoder-decoder}
\end{figure*}
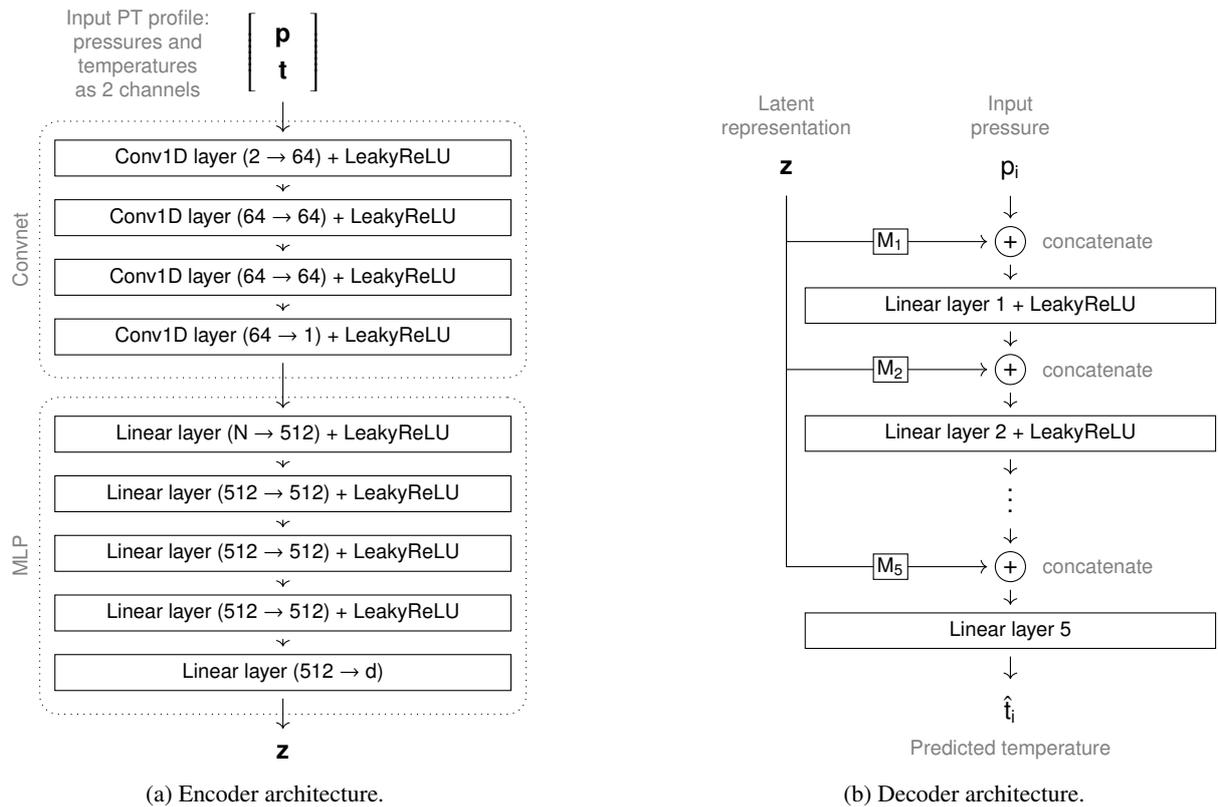

Perhaps not surprisingly, when experimenting with different encoder and decoder network architectures, we found that there does not seem to exist a \enquote{one-size-fits-all} configuration that consistently yields the best results across all data sets and latent space dimensions.
However, keeping in mind the proof-of-concept nature of this work, we decided to report only a single, simple experimental setup that gave reasonable results in all cases.
\Cref{fig:encoder-decoder} shows a visual illustration of our choices.
For practical applications, where one is considering a specific data set and latent size, one can fine-tune the configuration to that particular setting.

Our encoder is a standard convolutional neural network (CNN), consisting of 4 convolutional layers followed by a multi-layer perceptron (MLP) with 5 fully-connected layers.
It takes as input a PT profile where $\bm{p}$ and $\bm{t}$ are given as 2 channels.
The convolutional layers have 64 channels, with the last layer returning only a single output channel.
The kernel size of all layers was chosen as 1.
The output of the convolutional part of the network is then passed to a multi-layer perceptron (MLP; i.e., a series of fully-connected layers) with 3 hidden layers and 512 units each, which outputs the latent representation $\z \in \R^d$.

The decoder is even simpler than the encoder, not least to ensure that the evaluation during a retrieval is fast.
It consists of a single MLP, with 3 hidden layers and 512 units.
The \enquote{conditioning $D$ on $\z$} part is achieved as follows: 
Each layer receives the output of the previous layer concatenated with $M_i \z$ as its input, where the $M_i \in \R^{m \times d}$ are learnable matrices that map the $d$-dimensional $\z$ vector onto a vector in $\R^m$.
For this work, we have set $m = 16$.
We note that the number of layers and the activation function used for the decoder have an effect on which kind of functions (i.e., PT profiles) the model can learn. 
However, as real PT profiles are usually smooth, low-frequency functions with few inflection points, our simple choice seems to work well enough in practice.

All of our networks use LeakyReLU activation functions.
No further regularization (e.g., dropout or batch norm) was used.

\subsection{Implementation}
We implement our models in Python using \texttt{PyTorch}~\citep{Paszke_2019} and use the \texttt{PyTorch Lightning} wrapper\footnote{\url{https://lightning.ai/pytorch-lightning}} to run the training.
All of our code is available on GitHub.\footnote{\url{https://github.com/timothygebhard/ml4ptp}}

\subsection{Data split and pre-processing}
\label{par:data-split}
We split both the \pyatmos and the \goyal dataset randomly into a training and a test set.
For the \pyatmos data, the training set size is \num{100000}; the remaining \num{24314} are held out for the evaluation.
For the \goyal data, we use \num{10000} profiles for training and keep \num{1290} for testing.
During training, the training set is again split into two datasets: one that is actually used to train the model, and one validation set that we use to monitor the training progress.
We use a random 90\%\,/\,10\% split here, and run all experiments 3 times with different random seeds to test how much the final model depends on the exact training data and the initialization of the weights of the network.

We built our models to work with the decimal logarithm of the pressure rather than the pressure itself; otherwise, the scale of the pressure values would span several orders of magnitude, making it difficult for the neural net to learn anything. 
In addition, both (log)pressure and temperature are normalized using a standard score transform, where we subtract the mean and divide by the standard deviation (computed on the training set).

\subsection{Training}
We train our models for \num{1000} epochs using the \texttt{AdamW} optimizer \citep{Loshchilov_2017} with default values for the momentum and weight decay parameters.
The batch size is set to 512, and we modulate the learning rate using a \texttt{OneCycleLR} scheduler, with 5\% warm-up and a maximum learning rate of $\eta_\text{max}$ = \num{5e-4}.
Additionally, we use gradient clipping with a threshold of \num{1.0} on the gradient norm.
Again, these hyperparameters were not optimized.
Finally, to reduce overfitting, we only save the model with the lowest validation loss (\enquote{early stopping}).

We run our training on modern GPUs (NVIDIA V100 and A100).
Depending on the dataset, training a model to convergence usually takes between half an hour and two hours.

\subsection{Exporting and running trained models}
Trained models are exported using ONNX\footnote{\url{https://onnx.ai}}, an open-source format for ML interoperability.
Using the standard runtime engine\footnote{\url{https://onnxruntime.ai}}, evaluating our model for a given $\z$ on 100 pressure values takes about \SI{1.6}{\milli\second} on a standard laptop CPU (Intel Core i7).
If necessary, this can be further improved, for example by pruning or quantizing the model, or by using a specialized execution engine.

\subsection{Things that did (not) work}

To inform further research, we disclose a few ideas that we have tried and that did not work for us:

\begin{itemize}[leftmargin=*]
    \item \emph{Hypernet decoders:} In this setup, we used an additional neural network $H$ that takes in $\z$ and predicts the weights of the conditioned decoder network $D(\cdot \given \z)$.
    While this may seem like the most principled way of conditioning $D$ on $\z$, we found that in practice, this seems to give no advantage over the simpler decoder architectures described above.
    \item \emph{SIREN-style decoders:} We observed that when fitting only a single PT profile with a neural network, using sine activation functions (see \citealt{Sitzmann_2020}) leads to faster convergence and a smaller MSE than other activation functions.
    However, this advantage did not seem to uphold when learning a distribution over PT profiles.
    \item Batch normalization, stochastic weight averaging (SWA), and larger batch sizes all seemed to decrease the performance.
\end{itemize}

Conversely, two things that we did find to be important are:

\begin{itemize}[leftmargin=*]
    \item \emph{Normalization:} As mentioned before, we work with the logarithm of the pressure instead of the pressure itself, and normalize all PT profiles using a standard score transform.
    \item \emph{Initialization:} In some cases, the initial weights of the encoder produce representations $\z$ that are all very close to 0. 
    This usually causes the model to collapse and not learn anything.
    Our training procedure implements a sanity check for this case, which re-initializes the encoder weights until the mean norm of $\z$ is above some arbitrarily chosen threshold.
\end{itemize}

\section{Additional atmospheric retrievals}
\label{sec:additional-retrievals}

In \cref{subsec:retrieval-experiment}, we demonstrated the applicability of our method in atmospheric retrieval for a scenario where the true PT profile is taken from the distribution of PT profiles that we used to train our model.
Here, we consider a more challenging situation where the true PT profile is not covered by the training data.
In this case, we cannot expect the retrieval to recover the ground truth PT profile perfectly; however, we can show that our method still behaves reasonably and produces interpretable results.

To this end, we perform cloud-free atmospheric retrievals for two PT profiles of planets around a Sun-like star published in \citet{Rugheimer_2018}, representing (1)~a present-day Earth and (2)~a version of Earth \SI{0.8}{\giga\year} ago, around the time of the Neoproterozoic Oxygenation Event (NOE), which is associated with the emergence of large and complex multicellular organisms \citep[e.g.,][]{Knoll_2017}.%
\footnote{
    We obtained the PT profiles through private correspondence with the first author of \citet{Rugheimer_2018}.
}
Both profiles (and the corresponding spectra) were obtained using \mbox{EXO-Prime} \citep{Kaltenegger_2010}, a self-consistent coupled one-dimensional code developed for terrestrial exoplanets, which is different from the code that was used to generate our training data (ATMOS; see \cref{sec:datasets}).
\mbox{EXO-Prime} uses 1D models for the climate \citep{Kasting_1986,Pavlov_2000,HaqqMisra_2008}, the photochemistry \citep{Pavlov_2002}, and the line-by-line radiative transfer \citep{Traub_1976,Kaltenegger_2009}.
The emission spectra published alongside these PT profiles in \citet{Rugheimer_2018} were obtained using non-constant (i.e., pressure-dependent) abundance profiles for the different species.
While this is obviously a more realistic approach, running our atmospheric retrieval procedure with pressure-dependent abundances would complicate things without providing more insight into our method.
Therefore, to keep things simple, we have recalculated the spectra using petitRADTRANS \citep{Molliere_2019} assuming constant abundance profiles (using the ground truth values from \cref{tab:retrieval-parameters}), and used these as targets for the atmospheric retrieval procedure.
While this is, as noted above, a simplification compared to the self-consistent spectra of \citet{Rugheimer_2018}, recomputing the spectra with petitRADTRANS (which we use for retrieval) has the added benefit of reducing the potential bias from comparing spectra computed with different codes or opacity tables (cf., e.g., \citealt{Alei_2022a}), which is beyond the scope of this work.
Finally, the general setup (i.e., tools, parameters, priors, etc.) for the atmospheric retrievals remains the same as in \cref{subsec:retrieval-experiment}.

\begin{figure*}
    \centering
    \begin{subfigure}[t]{90mm}
        \centering
        \includegraphics[width=87mm]{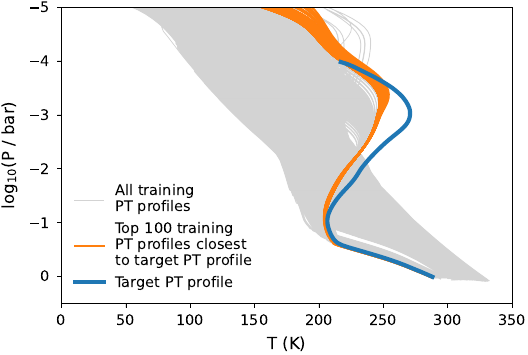}
        \newline
        \vspace{5mm}
        \includegraphics[width=87mm]{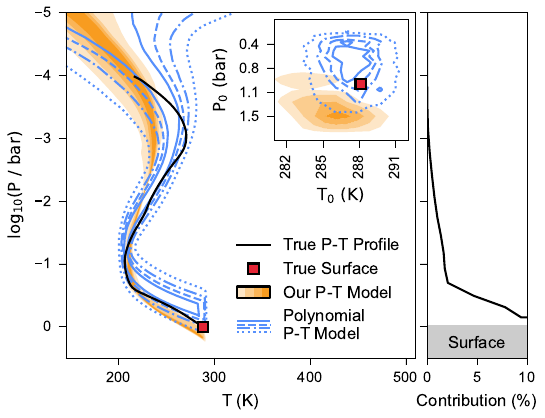}
        \newline
        \vspace{5mm}
        \includegraphics[width=87mm]{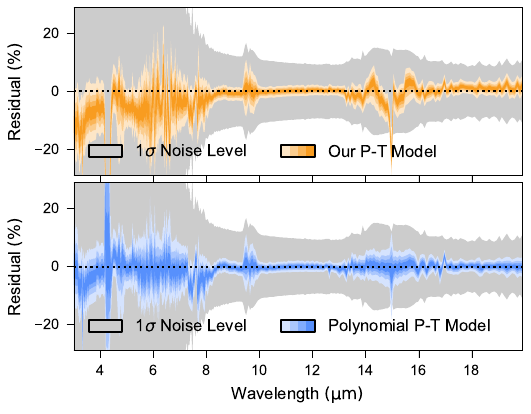}
        \subcaption{Results for the Modern Earth case.}
    \end{subfigure}
    \hfill
    \begin{subfigure}[t]{90mm}
        \centering
        \includegraphics[width=87mm]{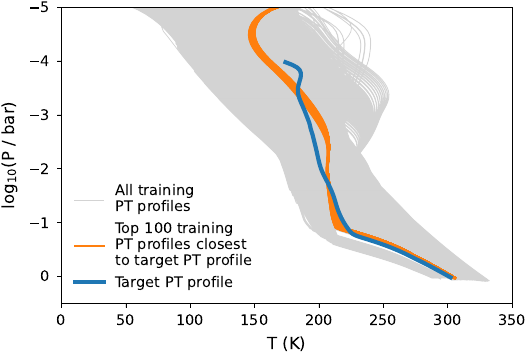}
        \newline
        \vspace{5mm}
        \includegraphics[width=87mm]{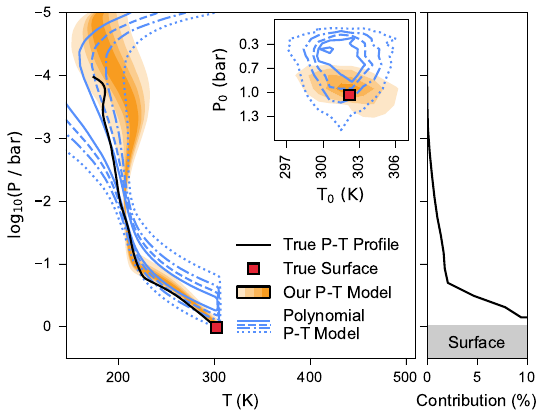}
        \newline
        \vspace{5mm}
        \includegraphics[width=87mm]{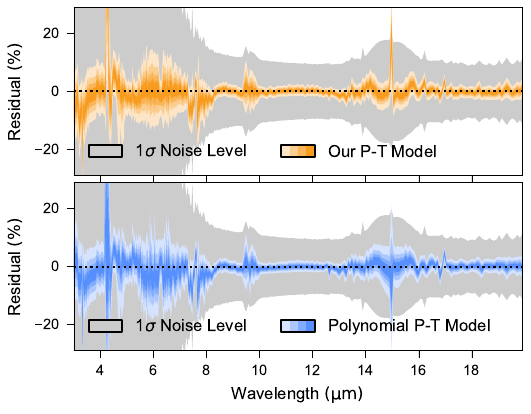}
        \subcaption{Results for the NOE Earth case.}
    \end{subfigure}
    \caption{
        Retrieval results for both the Modern Earth and the NOE Earth case.
        The top panel in each subfigure shows the respective target PT profile together with the \num{100} best-matching PT profiles from the training set.
        (\enquote{Best-matching} here refers to the unweighted mean squared difference between the PT profiles, where the average is taken over all atmospheric layers.)
        The middle panel shows the PT profiles obtained by our simulated atmospheric retrieval, both for our model and for the polynomial baseline.
        Finally, the bottom panel shows the respective fitting errors on the spectrum; cf. the right side of \cref{fig:retrieval-results}.
    }
    \label{fig:me-noe-results}
\end{figure*}

We present the results for both the Modern Earth case and the NOE Earth case in \cref{fig:me-noe-results}.
Looking at these results, we find that---as expected---the retrieval routine is unable to fit the respective ground truth PT profile perfectly.
This is because these PT profiles were not part of the training set (as illustrated by the top panel in \cref{fig:me-noe-results}) and thus the model did not learn to produce such PT profiles.
However, if we look at the retrieved PT profiles in the middle panel of \cref{fig:me-noe-results}, we see that they match well with the set of PT profiles in the training data that are closest to the respective target profile.
(\enquote{Closest} here refers to the smallest mean squared difference when comparing temperatures over the same grid of pressure values.)
This behavior---recovering the best-matching PT profiles from the training set---is the best outcome one can hope for in cases where the model is confronted with data that is has never encountered before.
After all, one of the advantages of our method is that it allows using a given grid of PT profiles as a prior for an atmospheric retrieval (in a parameterized, continuous fashion). 
A model that could \enquote{generalize out of the distribution} and also fit PT profiles that it never encountered during training would run counter to this idea.

\section{Additional result plots}

\begin{figure*}[p]
    \centering
    \labelbox{2.5mm}{2.5mm}{0}{}\hfill%
    \labelbox{40mm}{2.5mm}{0}{\hspace{8mm}dim(z) = 1}\hfill%
    \labelbox{40mm}{2.5mm}{0}{\hspace{8mm}dim(z) = 2}\hfill%
    \labelbox{40mm}{2.5mm}{0}{\hspace{8mm}dim(z) = 3}\hfill%
    \labelbox{40mm}{2.5mm}{0}{\hspace{8mm}dim(z) = 4}\\
    \labelbox{2.5mm}{36mm}{90}{\hspace{6mm}Best RMSE}\hfill%
    \includegraphics{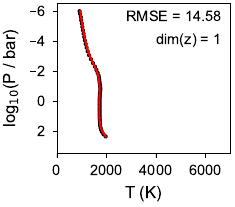}\hfill%
    \includegraphics{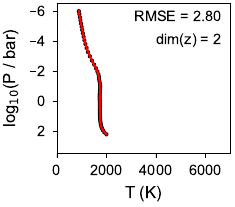}\hfill%
    \includegraphics{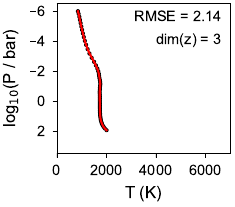}\hfill%
    \includegraphics{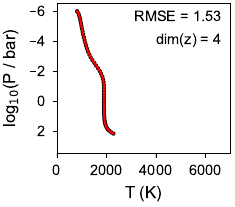}\\[2.5mm]
    \labelbox{2.5mm}{36mm}{90}{\hspace{6mm}Median RMSE}\hfill%
    \includegraphics{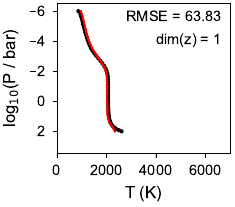}\hfill%
    \includegraphics{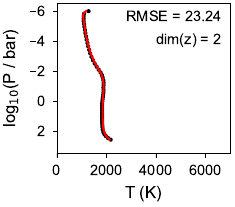}\hfill%
    \includegraphics{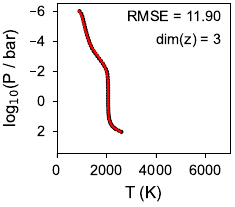}\hfill%
    \includegraphics{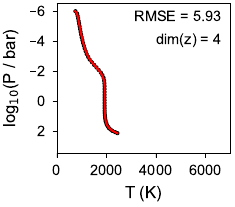}\\[2.5mm]
    \labelbox{2.5mm}{36mm}{90}{\hspace{6mm}Worst RMSE}\hfill%
    \includegraphics{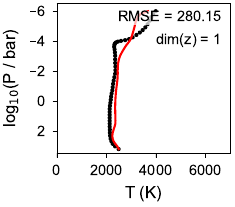}\hfill%
    \includegraphics{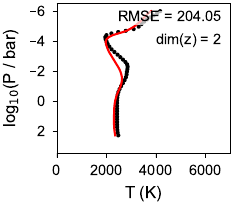}\hfill%
    \includegraphics{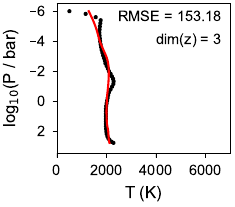}\hfill%
    \includegraphics{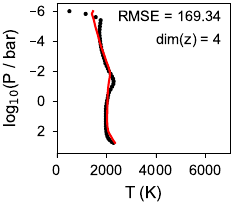}
    \caption{
        Same as \cref{fig:example-profiles-pyatmos} (i.e., PT profiles from our test set, and the best fit with our model), but for the \goyal dataset.~%
        \codelink{https://github.com/timothygebhard/ml4ptp/blob/main/scripts/plotting/fig-4-example-pt-profiles/plot-pt-profile.py}
    }
    \label{fig:example-profiles-goyal-2020}
\end{figure*}

\begin{figure*}[t]
    \centering
    \includegraphics{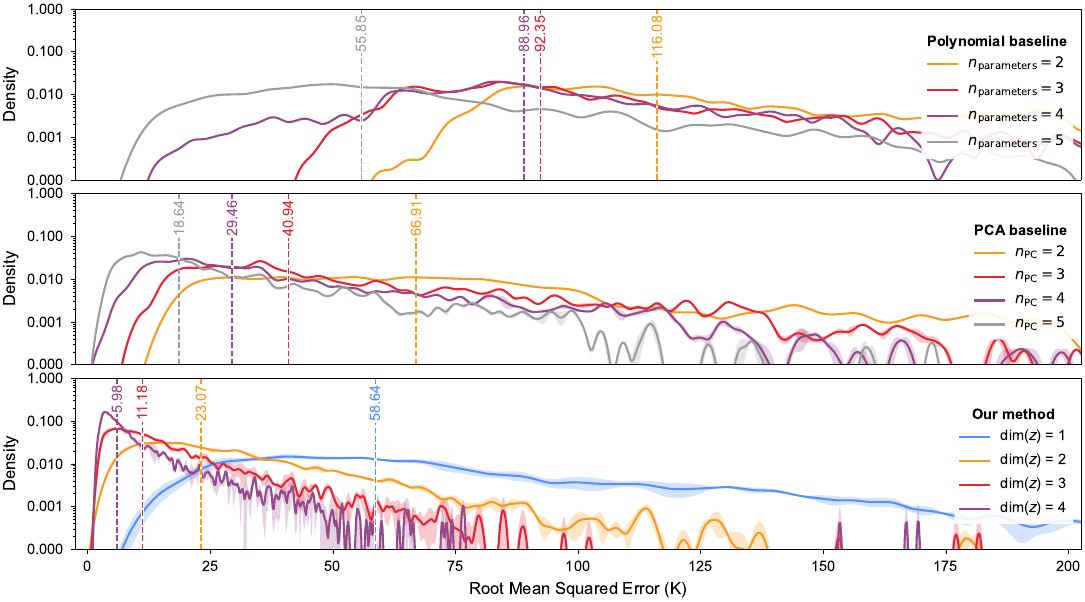}    
    \caption{
        Same as \cref{fig:error-distribution-pyatmos} (i.e., a comparison of the fitting error distribution on the test set), but for the \goyal dataset.~%
        \codelink{https://github.com/timothygebhard/ml4ptp/blob/main/scripts/plotting/fig-5-error-distributions/plot-error-distributions.py}
    }
    \label{fig:error-distribution-goyal-2020}
\end{figure*}

\begin{figure*}[t]
    \centering
    \includegraphics{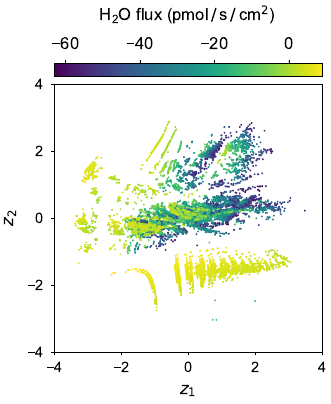}\hfill%
    \includegraphics{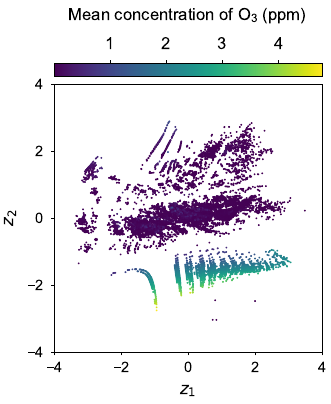}\hfill%
    \includegraphics{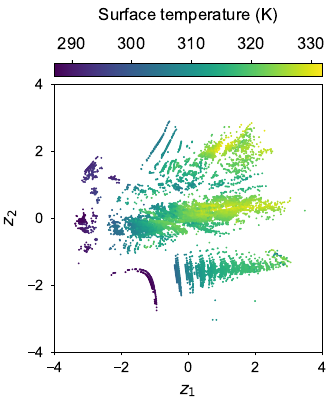}
    \caption{
        Same as \cref{fig:color-coded-z-goyal-2020} (i.e., latent representations $\z$ color-coded with atmospheric properties), but for the \pyatmos dataset.~%
        \codelink{https://github.com/timothygebhard/ml4ptp/blob/main/scripts/plotting/fig-8-color-coded-z/plot-color-coded-z.py}
    }
    \label{fig:color-coded-z-pyatmos}
\end{figure*}

In \cref{fig:example-profiles-goyal-2020}, \cref{fig:error-distribution-goyal-2020} and \cref{fig:color-coded-z-pyatmos}, we are showing additional result plots for the experiments described in the main text.

\end{document}